\documentclass{aa}
\usepackage{graphicx}
\usepackage{txfonts}
\usepackage{fixltx2e}

\usepackage{multirow}
\usepackage{natbib}
\usepackage{placeins}
\usepackage{url}
\usepackage{relsize}

\newcommand{\xmm}{{XMM-{\em Newton} }}
\newcommand{\xmmns}{{XMM-{\em Newton}}}
\newcommand{\swift}{{\em Swift }}
\newcommand{\swiftns}{{\em Swift}}
\newcommand{\rosat}{{\em ROSAT }}
\newcommand{\srcnamelong}{{XMMSL1~J074008.2-853927 }}
\newcommand{\srcnamelongns}{{XMMSL1~J074008.2-853927}}
\newcommand{\srcname}{{XMMSL1~J0740-85 }}
\newcommand{\srcnamens}{{XMMSL1~J0740-85}}
\newcommand{\srcnamemlong}{{2MASX~07400785-8539307 }}
\newcommand{\srcnamemlongns}{{2MASX~07400785-8539307}}

\newcommand{\swtd}{{SWIFT~J164449.3+573451 }}
\newcommand{\swtdns}{{SWIFT~J164449.3+573451}}
\newcommand{\assasn}{{ASSASN-14li }}
\newcommand{\assasns}{{ASSASN-14li}}
\newcommand{\fluxUnits}{{ergs s$^{-1}$cm$^{-2}$ }}
\newcommand{\fluxUnitsns}{{ergs s$^{-1}$cm$^{-2}$}}
\newcommand{\lumUnits}{{ergs s$^{-1}$ }}
\newcommand{\lumUnitsns}{{ergs s$^{-1}$}}

\newcommand{\chired}{{$\chi^{2}_{r}$}}
\newcommand{\msolar}{$M_{\sun}$ }
\newcommand{\msolarns}{$M_{\sun}$}

\begin{document}
   \title{\srcnamelongns: a tidal disruption event with thermal and non-thermal components}

   \subtitle{}

   \author{R.D. Saxton
          \inst{1}
          \and
          A.M. Read\inst{2}
          \and
          S. Komossa\inst{3,4}
          \and
          P. Lira\inst{5}
          \and
          K.D. Alexander\inst{6}
          \and
          M.H. Wieringa\inst{7}
          }

   \offprints{R. Saxton}

   \institute{XMM SOC, ESAC, Apartado 78, 28691 Villanueva de la Ca\~{n}ada, Madrid
              , Spain\\
              \email{richard.saxton@sciops.esa.int}
         \and
             Dept. of Physics and Astronomy, University of Leicester, Leicester LE1 7RH, U.K.
         \and
             QianNan Normal University for Nationalities, Longshan Street, Duyun City of Guizhou Province, China
         \and
             Max Planck Institut f\"ur Radioastronomie, Auf dem Huegel 69, 53121 Bonn, Germany
         \and
             Universidad de Chile, Observatorio Astronomico Nacional Cerro Calan, Santiago, Chile
         \and
             Harvard-Smithsonian Center for Astrophysics, 60 Garden St., Cambridge, MA 02138, USA
         \and
             Australia Telescope National Facility, CSIRO Astronomy and Space Science, PO box 76, Epping, NSW 1710, Australia
        }

   \date{Received September 15, 1996; accepted March 16, 1997}

  \abstract
  % context heading (optional)
  {} 
  % aims heading (mandatory)
   {We study X-ray bright tidal disruption events (TDE), close 
to the peak of their emission, with the intention of understanding 
the evolution of their light curves and spectra.}
  % methods heading (mandatory)
   {Candidate TDE are identified by searching for soft X-ray flares from non-active galaxies in recent \xmm slew data.}
  % results heading (mandatory)
   {In April 2014, X-ray emission was detected
from the galaxy \srcnamelong (a.k.a. \srcnamemlongns), a factor 20 times higher than an upper limit
from 20 years earlier. Both the X-ray and UV flux subsequently fell, by factors of 70 and 12 respectively. 
 The bolometric luminosity peaked at L$_{bol}\sim2\times10^{44}$ \lumUnits
with a spectrum that may be modelled with thermal emission in the UV band,
a power-law with $\Gamma\sim2$ dominating in the X-ray band above 2 keV
and a soft X-ray excess with an effective temperature of $\sim86$ eV.
Rapid variability locates the X-ray emission to within $<$73 $R_{g}$ of the nuclear black hole.
Radio emission of flux density $\sim 1$ mJy, peaking at 1.5 GHz was 
detected 21 months after discovery.
Optical spectra indicate that the galaxy, at a distance 
of 73 Mpc (z=0.0173), underwent a starburst 2 Gyr ago and is now quiescent.
We consider a tidal disruption event to be the most likely cause of the flare.
If this proves to be correct then this is a very clean example of a disruption
exhibiting both thermal and non-thermal radiation.}
   {}


   \titlerunning{\srcnamelong : a tidal disruption event}

   \maketitle
%
%________________________________________________________________

\section{Introduction}

When a star approaches very close to a supermassive black hole (SMBH) the gravitational shear
can overcome the self-gravitation of the star causing it to be torn apart \citep{Hills} in what
is known as a tidal disruption event (TDE).
About half of the stellar debris remains gravitationally bound to the hole and is accreted 
on return, causing a flare of thermal
radiation which peaks in the EUV band \citep{Rees88}. This radiation signature has been
observed in the soft X-ray \citep{Komossa1242,Bade96,Komossa99b,Esquej07,
Saxton12,Maksym10,Lin15} and UV \citep{Gezari06,Gezari08,Gezari09} bands.
Optical flares have also been detected \citep{vanVelzen11,Cenko12b,Gezari12,Arcavi14} from candidate TDE, with an apparent temperature of
a few $\times10^{4}$ K, cooler than that expected from an accretion disc \citep[e.g. ][]{Bonning07}.
Current models for the optical emission centre on a thick screen of debris, which 
reprocesses the inner thermal radiation and re-emits it in the optical band \citep{Strubbe09,Strubbe11,GuillochonPSJ}. 
\citet{Strubbe15} suggest that the initial accretion phase will be highly super-Eddington, resulting in a
massive radiation-driven outflow of material that will absorb the radiation from the central engine
and convert it into optical photons. At a later stage, the screen density will drop and the inner
thermal radiation will become visible. A further signature of reprocessing 
is given by broad,
low-ionisation optical  lines which indicate a large solid angle of 
absorbing material orbiting the nucleus. \citet{Komossa08} discovered 
super-strong broad Hydrogen, Helium and highly
ionised narrow Iron lines which faded over time in SDSS~J095209.56+214313.3. 
Bright fading optical broad lines were also seen in other archival SDSS 
data \citep{Wang11} and in \assasn \citep{Holoien16a}. Broad Nitrogen 
and Helium lines were also detected in the UV band from \assasn 
\citep{Cenko16}.
The rate at which galaxies produce a soft X-ray TDE has been estimated from \rosat observations as $9\times10^{-6}$ events/galaxy/year \citep{Donley}; 
from \xmm slew
observations as $2\times10^{-4}$ events/galaxy/year \citep{Esquej08} and from
galaxy clusters $1.2\times10^{-4}$ events/galaxy/year \citep{Maksym10}.
These numbers are comparable to the rate of TDE, detected from 
their peak
optical emission of $2\times10^{-5}$ events/galaxy/year \citep{vvFar14}.
Even though these estimates are based on small numbers of events, they
are difficult to reconcile with a model where the majority of X-ray events are
invisible due to a thick absorber. Furthermore, strong two-temperature 
soft X-ray and optical emission 
has been observed simultaneously from the well-monitored TDE, \assasns, 
implying that
our view to the centre is rather unobscured, apart from a highly-ionised, slowly-expanding,
screen of material with a column of $\sim10^{23}$ cm$^{-2}$ \citep{Miller,Cenko16}.
Finally, the exceptional X-ray softness of the first identified TDEs strongly argues against large amounts of obscuring material \citep[e.g. ][]{KomossaBade}.

Radio emission has been associated with TDE in several events. \swtd exhibited
strong, variable radio emission compatible with a forward shock from material swept-up 
by an on-axis relativistic jet \citep{Zauderer11}. This was also seen in the more distant
object SWIFT~2058+0516 \citep{Cenko12b}. Both of these were discovered from their super-Eddington,
{\em hard} ($>2$ keV)  X-ray emission by \swiftns, as was a further jetted event, SWIFT~1112-8238
\citep{Brown15}. Hard X-ray emission dominated the
spectrum of NGC~5485 \citep{Walter13}, also accompanied by  
radio emission 
\citep{Irwin} which led to the suggestion that this was another relativistic event, this time
viewed off-axis \citep{Lei16}. Transient radio emission was seen in \assasn 
\citep{Alexander16,vanVelzen16} but at a level which may be consistent with a non-relativistic outflow rather than a jet. 
It is an open, and interesting, question as to why these events
should show a hard X-ray component, compatible with power-law emission, while 
many TDE show only thermal X-ray emission 
\citep[e.g.][]{Komossa02, Esquej07, Saxton12, Maksym10,Lin15,Holoien16a}.  
In 2MASS 0619-65 \citep{Saxton14} hard X-ray emission was not
accompanied by a radio flare and it is not yet decided whether 
the high-energy photons, which are seen in TDE come
from a jet or from the hot, comptonising, electron plasma which is believed
to be responsible for the same component in AGN \citep{SunTit, HarMar}.
Even in \swtdns, evidence for X-ray reprocessing within a few gravitational
radii ($R_{g}$) of the black hole has called into question the hitherto accepted 
view that the huge X-ray luminosity ($L_{X}\sim10^{48}$\lumUnits) 
was produced by a jet \citep{Kara16}. 

In April 2014, the \xmm slew survey \citep{Saxton08} detected a flare from the nucleus of
the quiescent galaxy \srcnamelong which further monitoring showed
to be both thermal and non-thermal, having simultaneous flares in the UV, soft X-ray and hard X-ray bands. In Section 2 we discuss the discovery of this flare and the source identification;
in Sections 3, 4 \& 5 we present X-ray, UV, optical and radio follow-up 
observations; in Section 6 we perform a temporal and spectral analysis
of the source and in Section 7 we discuss the flare
characteristics within the TDE model. The paper is summarised in Section 8.

A $\Lambda$CDM cosmology with ($\Omega_{M},\Omega_{\Lambda}$) = (0.27,0.73)
and  $H_{0}$=70 km$^{-1}$s$^{-1}$ Mpc$^{-1}$ has been assumed throughout.

\section{X-ray flare identification}
\label{sec:ident}
During the slew 9262100003, performed on April 1$^{st}$ 2014, \xmm 
\citep{jansen} detected
a source, \srcnamelong (hereafter \srcnamens), with an EPIC-pn, medium filter,
 0.2--2 keV count rate of 
$4.2\pm0.6$ count s$^{-1}$. 
A crude analysis may be performed on the 45 photons in the 
slew spectrum to investigate the gross spectral properties of the 
detection. Detector matrices were calculated, taking into account 
the transit of the
source across the detector, using a technique
outlined in \citet{Read08}. A simple absorbed power-law fit gives a slope
$\Gamma=2.9\pm{0.5}$ assuming no intrinsic
absorption above the Galactic value 
of $1.5\times10^{21}$cm$^{-2}$ \citep{Willingale}.
This corresponds to an absorbed flux of 
$F_{0.2-2}\sim4.9\pm{0.7}\times10^{-12}$ \fluxUnits using the above model.
The count rate in the 2--10 keV band is $0.8\pm0.3$ count s$^{-1}$. 
Using the best spectral model from follow-up observations this gives 
 a flux, $F_{2-10}\sim4.8\pm{1.8}\times10^{-12}$ \fluxUnitsns.  
We calculate a 2-sigma upper limit from the \rosat All-Sky Survey 
(RASS) 
 at this position of 0.021 count s$^{-1}$ \citep[see][for a description of the upper limit
calculation]{Esquej07}; a factor 20 lower 0.2--2 keV flux using the same spectral model. 

The source position lies $3.8\arcsec$ from the galaxy 
\srcnamemlong \citep[z=0.0173;][]{Huchra12} which with 
J=$11.92\pm{0.03}$, H=$11.15\pm{0.04}$, K=$10.96\pm{0.05}$, R=10.7 is the 
only bright optical and infrared source within the $8\arcsec$ error circle 
\citep{Saxton08}.

%In archival data it was found that this position had been observed in four
% earlier slews yielding 0.2--2 keV count rates of 
%$<0.37$ count s$^{-1}$ in 2007-07-07, $<0.75$ count s$^{-1}$ in 2011-10-13,
%$1.0\pm0.2$ count s$^{-1}$ from 2012-06-01 and $<0.67$ count s$^{-1}$ in 
%2012-06-22.

\section{X-ray and UV observations}
\label{sec:xuvobs}
\begin{figure}
\centering
%\rotatebox{0}{\includegraphics[height=6.5cm]{0740_SlewSwiftXRTpos_on_2MASS.eps}}
%\rotatebox{0}{\includegraphics[height=6.5cm]{slew_0740_4.eps}}
\rotatebox{0}{\includegraphics[height=6.5cm]{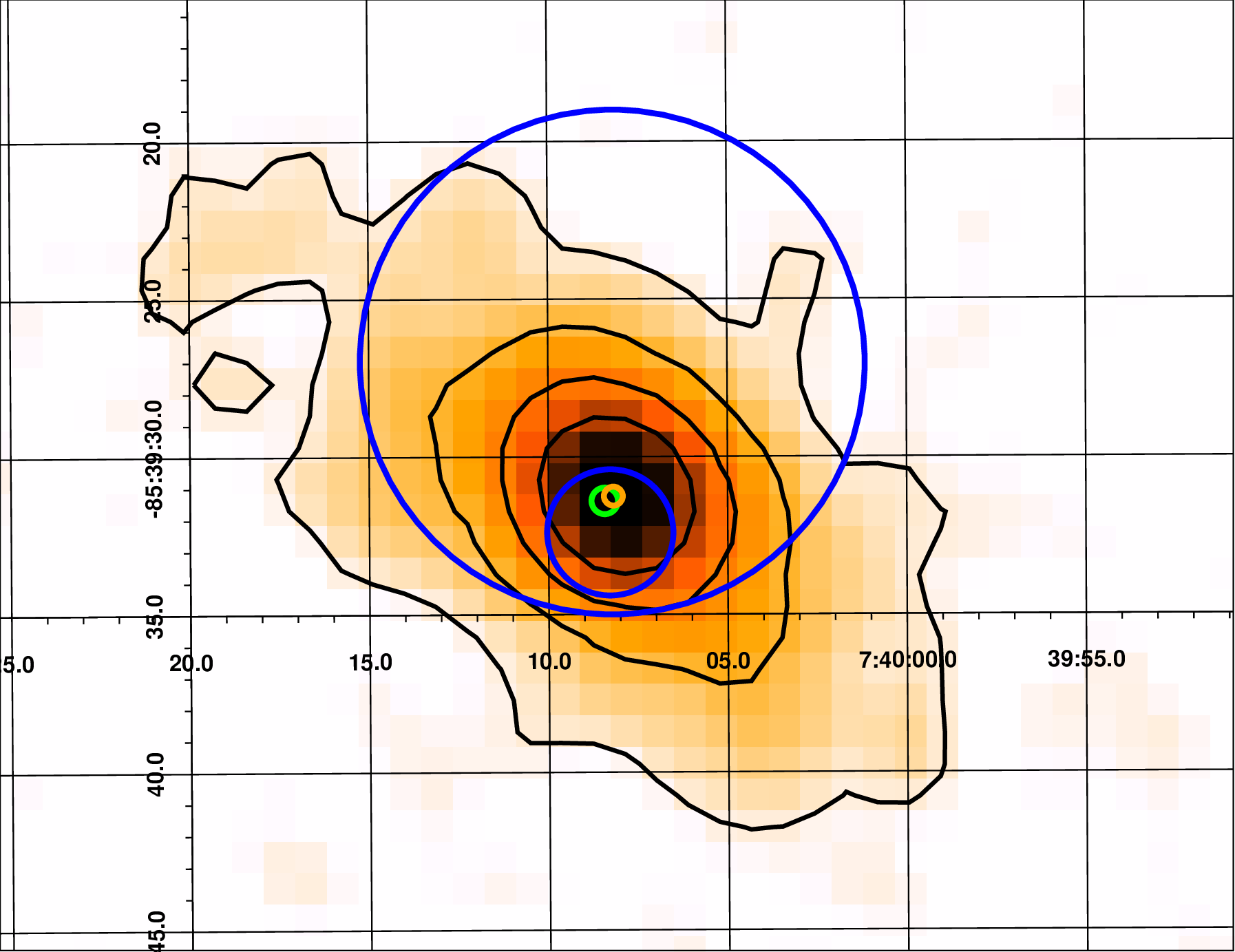}}
\caption[dss image]
{ \label{fig:swiftimage} A 2MASS image of the galaxy, 
 shown with the \xmm slew (blue, 1-$\sigma$, 8\arcsec radius), UVOT-enhanced
\swift-XRT (blue, 90\% conf., 2.0\arcsec radius; see text),
\swift-UVOT (green, 90\% conf., 0.42\arcsec radius) 
and ATCA, radio (orange, 1-$\sigma$, 0.3\arcsec radius) error circles, 
centred on the detections.
}
\end{figure}

An X-ray monitoring program was initiated with \swift \citep{Gehrels04} 
to 
follow the evolution of the source flux and spectrum. Snapshot 3ks observations
were made, initially once a week and then less frequently, with the \swiftns-XRT 
\citep{Burrows05} in photon
counting mode and the UV optical telescope \citep[UVOT;][]{Roming}.
 The \swiftns-XRT observations have been analysed following the 
procedure outlined in \citet{Evans} and the UVOT data have been 
reduced as described in \citet{Poole}. 
An accurate position for the source in the \swiftns-XRT field can be determined
by matching the UVOT field of view with the USNO-B1 catalogue and registering
the XRT field accordingly \citep{Goad}. The resulting source position, 
$\alpha_{J2000}$=07:40:08.28, $\delta_{J2000}$=-85:39:32.4 
 ($\pm 2.0\arcsec$; 90\% confidence)
is coincident with the galactic nucleus (see Fig.~\ref{fig:swiftimage}).
A more precise position is available from the UVOT images. 
The source position from the first \swiftns-UVOT observation, which used the
uvw2 filter, is
%where the light was completely dominated by the flare, 
$\alpha_{J2000}$=07:40:08.43, $\delta_{J2000}$=-85:39:31.4
with an error dominated by a systematic offset of $\pm 0.42\arcsec$
\citep[90\% confidence;][]{breeveld10}. 
This is consistent, within the error circle, with the 
galactic centre given by the 2MASS point source catalogue
($\alpha_{J2000}$=07:40:08.09, $\delta_{J2000}$=-85:39:31.3).
We can conclude that the flare was consistent with being from the nucleus of 
\srcnamemlong to within the accuracy of the UVOT pointing (150 pc at the
galaxy redshift).

In parallel,  \xmm pointed observations were triggered
on 2014-04-30 (due to scheduling issues this was split into three parts; 
obsid=0740340301, 0740340401, 0740340501) and 2015-01-12
(obsid=0740340601).  A summary of observations and exposure times is given in Tab~\ref{tab:xobs}.
In each observation, the EPIC-pn and MOS-1 cameras 
were operated in full frame mode with the {\em thin1} filter in place,
while the MOS-2 camera was used in small window mode with the medium 
filter. The source was too
faint for statistically significant data to be collected from the
reflection grating spectrometers. 
 
The XMM data were analysed with the \xmm Science Analysis System 
\citep[SAS v14.0.0;][]{Gabriel}. Light curves were extracted from the 
observations and searched for periods of high background flaring.
These periods were excluded from the datasets leaving for the first observation 
a total of 26.6 ks of useful data and for the second observation 16.1 ks.

%
% Following fluxes have been calculated from XSPEC using a
% model of NH=4.4E20, slope=2. Swift data fitted in range 0.3-2 keV,
% XMM in range 0.2-2 keV.
% Count rates have been removed from the original table which can 
% be found at the end of this file
%
\begin{center}
\begin{table*}[ht]
%\begin{table}
{\small
\caption{X-ray observation log of \srcname}
\label{tab:xobs}      % is used to refer this table in the text
\hfill{}
\begin{tabular}{l c l l l l l l l}
\hline\hline                 % inserts double horizontal lines
Mission$^{a}$ & Date & Exp time$^{b}$ & Flux$^{c}$ & uvw2 & uvm2 & uvw1 & U & B \\
              &      &   (s)    &    ($10^{-12}$\fluxUnitsns)  & mag & mag & mag & mag & mag \\
\\
ROSAT  & 1990  & 368  &  $<0.25$  & & & & & \\
XMM slew  &   2012-08-01  & 3.9  & $<0.7$  & & & & &  \\
XMM slew  &   2014-03-24  & 2.0  & $<1.7$  & & & & &  \\
XMM slew  &   2014-04-01  & 10.3  & $4.9\pm{0.7}$  & & & & &  \\
Swift  & 2014-04-14       &  2435   & $3.2\pm{0.21}$ & $16.35\pm0.03$ & & & & \\ 
Swift  & 2014-04-22       &  2697   & $0.43\pm{0.21}$ &$16.49\pm0.05$ & & & &  \\
XMM pointed & 2014-04-30  &  26600  & $0.5\pm{0.15}$ & & & $16.59\pm0.05$ & $15.75\pm0.01$& $15.52\pm0.01$ \\
Swift & 2014-05-05      &  1895   & $0.41\pm{0.12}$ & $16.80\pm0.10$& $16.94\pm0.08$&  $16.63\pm0.10$& & \\
Swift & 2014-05-12       &  2028  & $1.4\pm{0.12}$ & $17.07\pm0.10$& $17.11\pm0.10$&  $16.66\pm0.03$& & \\
Swift  & 2014-05-19       &  1399  & $0.36\pm{0.12}$ & $17.14\pm0.10$ & $17.23\pm0.20$& $16.76\pm0.03$ & &  \\
Swift  & 2014-05-26       &  509   & $0.35\pm{0.22}$ & &  & & & \\
Swift & 2014-05-31       &  1354  & $1.1\pm{0.12}$ & $17.10\pm0.06$&$17.22\pm0.07$ &  $16.68\pm0.04$& & \\
Swift  & 2014-07-04       &  954   & $0.40\pm{0.15}$ & $17.51\pm0.3$ & $17.52\pm0.10$ &  $17.24\pm0.40$& & \\
Swift  & 2014-08-01       &  2482  & $0.41\pm{0.12}$ & $17.52\pm0.08$&$17.77\pm0.10$ &  $16.99\pm0.03$ & &  \\
Swift  & 2014-08-29       &  1651  & $0.16\pm{0.12}$ & $17.28\pm0.10$ & $17.45\pm0.09$&  $16.96\pm0.04$& & \\
Swift  & 2014-09-26       &  2485  & $0.33\pm{0.12}$ & $17.40\pm0.05$& $17.56\pm0.09$&  $17.10\pm0.08$& & \\
Swift  & 2014-10-24       &  2073  & $0.20\pm{0.11}$ & $17.53\pm0.05$& $17.58\pm0.09$&  $17.01\pm0.04$& & \\
Swift  & 2014-11-21       &  137  & $0.30\pm{0.24}$ & $17.54\pm0.13$& & & & \\
Swift  & 2014-11-26       &  2472  & $0.21\pm{0.14}$ & $17.55\pm0.10$ & $17.72\pm0.15$&  $16.98\pm0.06$& & \\
XMM pointed & 2015-01-12  &  16100 & $0.30\pm{0.01}$ & & & $17.2\pm0.1$ & $15.92\pm0.01$& $15.60\pm0.01$ \\
Swift  & 2015-09-20       &  1890  & $<0.07$ & $19.0\pm0.3$ & $18.88\pm0.21$ &  $17.41\pm0.20$& & \\
\hline                        % inserts single horizontal line
\end{tabular}}
\hfill{}
\\
\\
$^{a}$ \xmmns, EPIC-pn camera: slew observations performed in {\em full frame}
mode with
the {\em medium} filter; pointed observations performed in {\em full frame} mode with
the {\em thin1} filter. \swiftns-XRT observations performed in {\em pc} mode. 
ROSAT sky survey data taken between 1990-08-12 and 1991-02-18.\\
$^{b}$ Exposure time in seconds. For the two XMM pointed observations, this is the
remaining exposure after removing times of high background flares. \\
$^{c}$ Absorbed flux in the 0.2-2 keV band, calculated by
fitting the spectrum of each observation with a two-component model of 
a power-law plus black-body
absorbed by a Galactic column of $1.5\times10^{21}$cm$^{-2}$. Upper limits
have been calculated using a power-law model of slope 2.9 absorbed by the
Galactic column.\\
\end{table*}
%\end{table}
\end{center}

\section{Optical observations}
\label{sec:optobs}

\begin{figure}
\begin{center}
\begin{minipage}{3in}
%\hspace*{-0.5in}    \rotatebox{0}{\includegraphics[height=7cm]{J0740_both_diff.eps}}
\hspace*{-0.5in}    \rotatebox{-90}{\includegraphics[height=8cm]{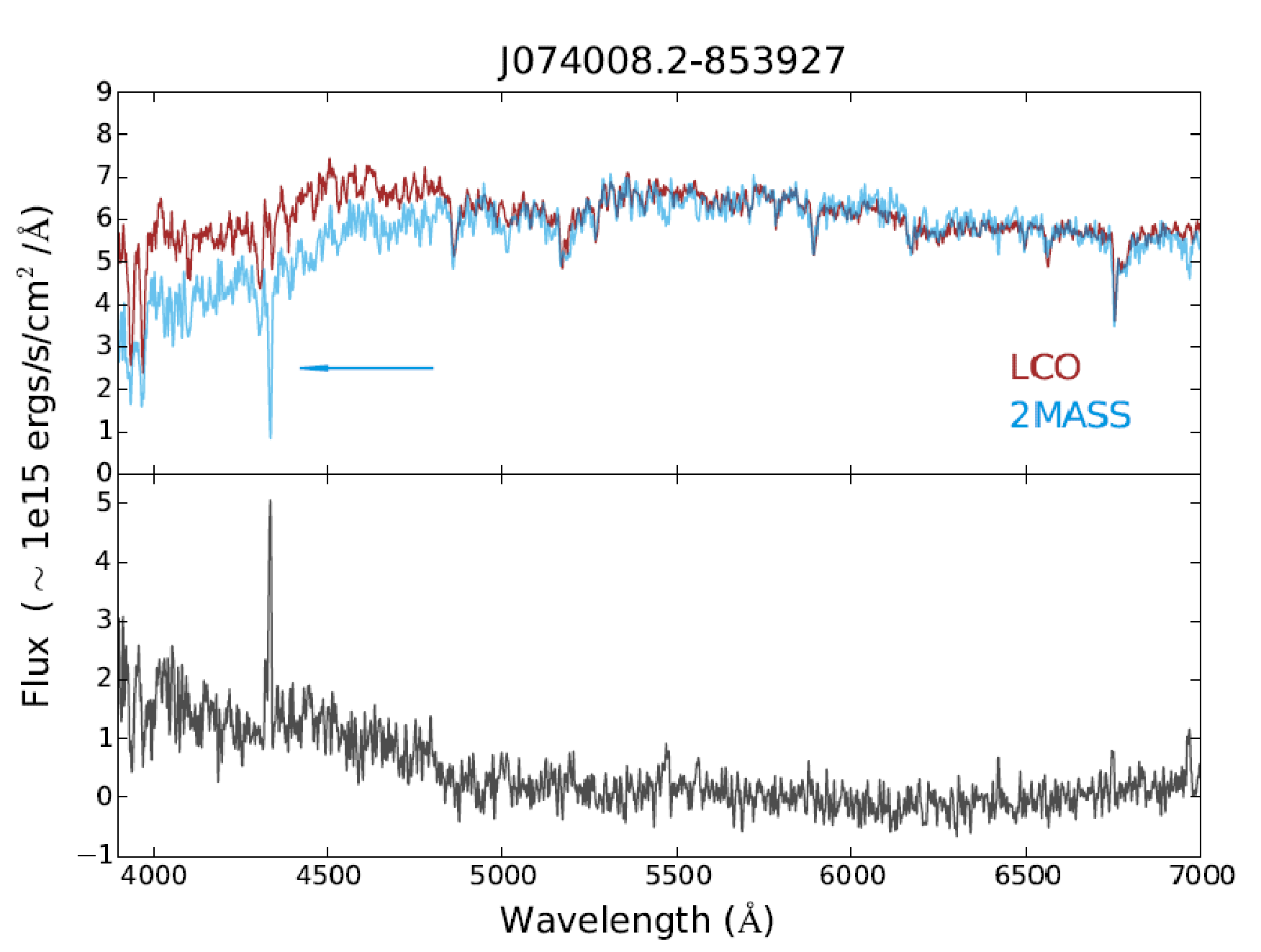}}
\end{minipage}

  \end{center}
\caption[\srcname Optical spectra]
{ \label{fig:optspec} Optical spectra of \srcname 
taken on 2002-01-17 with the CTIO 1.5m telescope for the
2MASS redshift survey (blue) and on 2014-05-05 with the Las Campanas,
du Pont 100-inch telescope (red). The CTIO spectrum has been 
scaled by a factor 1.55 to match the Las Campanas spectrum at the
red end. The lower panel shows the difference spectrum,
indicating a featureless excess in blue flux in the 2014 observation.
The apparent spike in the difference spectrum at $\sim4300\AA$ 
is caused by a spurious absorption feature in the CTIO spectrum.} 
\end{figure}

\srcname was observed by the CTIO 1.5m telescope, on 2002-01-17, as part of the 
2MASS redshift survey \citep{Huchra12}. The optical spectrum showed a non-active galaxy
(Fig.~\ref{fig:optspec}) at a redshift of $z=0.0173$.
We calibrated the raw spectrum using the spectrophotometric standards,
LTT 377 and LTT 3864. Observations at CTIO were obtained at 1.8 airmasses using a fixed E-W slit orientation (Macri, p. comm.), which might introduce slit losses at the blue end due to atmospheric differential refraction. However, for the given airmass, the refraction at 4000\AA\ with respect to the position of the image at 5000\AA\ corresponds to $\sim 0.95$ arcseconds \citep{Filippenko}. This is less than half the used slit width, and therefore differential refraction should not greatly impact the observation. 

We reobserved the source with the WFCCD imager-spectrograph
mounted on the du Pont 100-inch telescope located at Las Campanas
Observatory (LCO) in Chile on the 4th of May 2014. The spectrograph blue grism
was used resulting in a spectral resolution of $\sim 8 $\AA. Integration
time was 600 seconds. The data were reduced using IRAF packages in the
standard way. Wavelength calibration was achieved using HeArNe lamps,
while flux calibration was achieved using the observation of the LTT4364
spectrophotometric standard star. Even though the spectral range of the
observations extended to 9000 \AA, the analysis was conducted only in
the 3895-7150 \AA\ spectral window due to telluric absorption and heavy
fringing in the red end.

The red end of the spectrum obtained from LCO is 1.55 times brighter 
than that obtained from CTIO, which may be explained by the different 
slit widths of 4.8$\arcsec$ and 2.6$\arcsec$ 
 used respectively in each observation. 
In Fig.~\ref{fig:optspec} we compare the two calibrated spectra,
after scaling the CTIO spectrum to match the LCO observations,
and find that
the main difference is an increase in blue flux in the later observation 
below $\sim 4800\AA$. The difference spectrum is featureless apart from a
spurious absorption feature at $\sim4300\AA$ in the CTIO spectrum which
appears as a spike in the difference spectrum.

\begin{figure}
\begin{center}
\begin{minipage}{3in}
\hspace*{-0.4in}   \rotatebox{0}{\includegraphics[height=7cm]{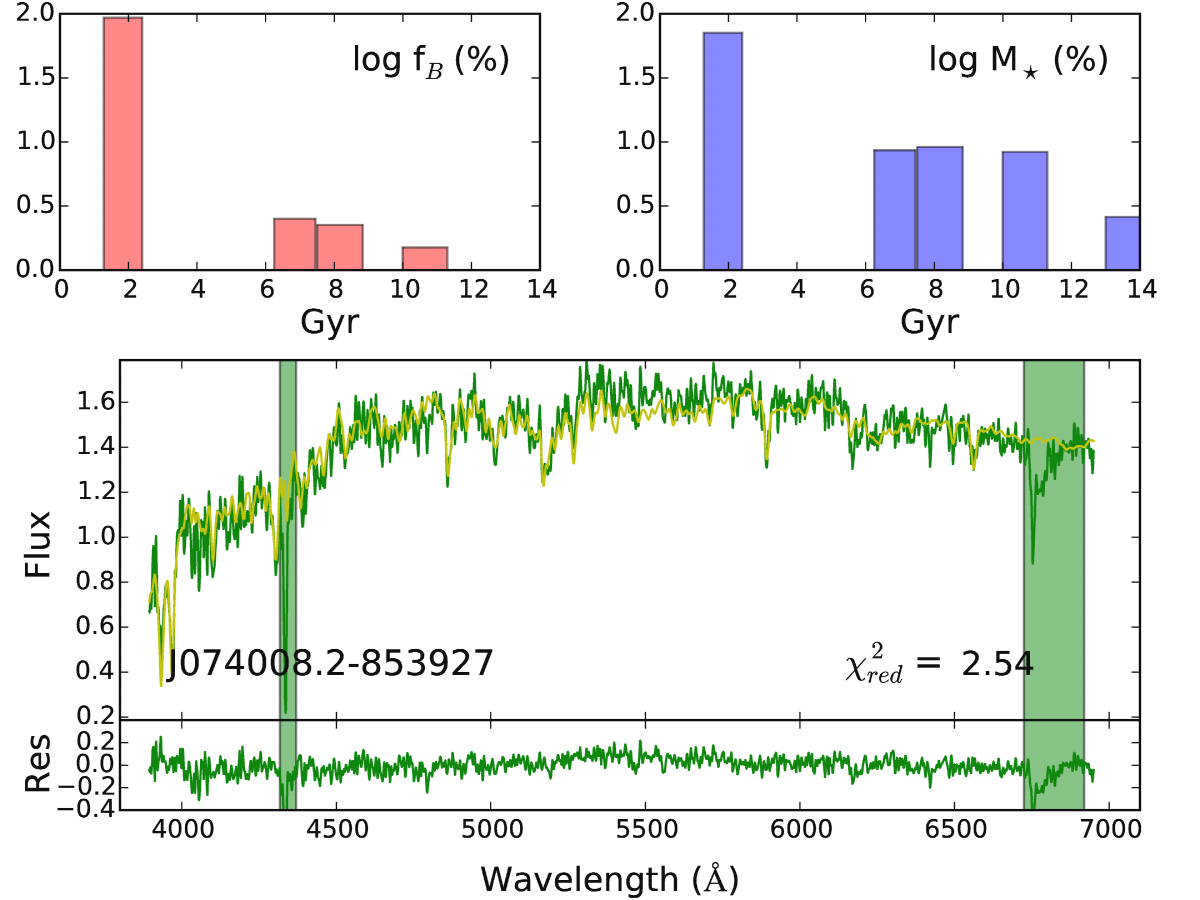}}
\end{minipage}

  \end{center}
\caption[\srcname Stellar population]
{ \label{fig:optspecfit} A stellar population fit of \srcname to the
2002-01-17 CTIO optical spectrum, with
the fitted age of stellar populations providing the
blue flux (upper-left) and the total mass (upper-right).
Residuals to the stellar fit are shown in the bottom panel.
See text for details of the fitting procedure.}
\end{figure}

The stellar population of \srcname was analysed using the
STARLIGHT spectral population synthesis code \citep{Fernand05,Fernand11}
on the CTIO spectrum. 
The code uses the \citet{Bruzual} single stellar
population models and requires the data to be corrected for foreground
Galactic extinction and taken to the rest frame. We used the \citet{Cardelli} 
extinction curve to correct for foreground reddening and
adopted $A_V=0.437$ and shifted the spectrum to the rest wavelength
using $z=0.017292$.
Figure~\ref{fig:optspecfit} presents the results from the 
spectral population synthesis
analysis. The bottom panel shows the observed and total synthetic
spectra and the fit residuals. The histograms present the percentage that
each of the single stellar populations provides to the current total
mass and B-band flux of the galaxy. 
The fit shows that the stellar emission
is dominated by a relatively young stellar population of age 2 Gyr but
that a significant amount of mass could be found in a much older
population.
The integrated absolute J magnitude of J0740 is -22.5, which is about 
1 magnitude below the knee of the J-band local galaxy luminosity 
\citep{Cole2001}. It is likely that, in such a massive galaxy, 
most of the stellar mass was formed at rather high redshifts, and this 
component would most likely be hidden under the light of the 
2 Gyr old component.

The residuals clearly show that no emission lines are present in
\srcname with a strict upper limit to the OIII emission line
flux of $4\times10^{-15}$ \lumUnitsns.

%%%%%%%%%%%%%%%%%%%%%%%%%%%%%%%%%%%%
%%%%%%%%%%%%%%%%%%%%%%%%%%%%%%%%%%%%

%%%%%%%%%%%%%%%%%%%%%%%%%%%%%%%%%%%%
%%%%%%%%%%%%%%%%%%%%%%%%%%%%%%%%%%%%

\section{Radio observation}
\label{sec:radioobs}

We observed the position of \srcname with the Australia Telescope Compact Array (ATCA) on 2015-11-14 UT 
in the 4cm band and 2015-12-01 UT in the 16cm and 15mm bands. We detected a source at 
$\alpha_{J2000}$=07:40:08.19, $\delta_{J2000}$=-85:39:31.25 ($\pm0.3\arcsec$), which is consistent with 
the \xmm and \swift positions. The source is detected at all frequencies, 
with a peak flux density 
of $1.19 \pm 0.06$ mJy at 1.53 GHz. At a distance of 73 Mpc, this corresponds to a 
luminosity $L_{\nu} = (7.7 \pm 0.4) \times 10^{27}$ erg s$^{-1}$ Hz$^{-1}$. This is slightly lower than 
the radio luminosity of the nearby thermal TDE ASSASN-14li \citep[likely produced by a non-relativistic 
outflow;][]{Alexander16,vanVelzen16}, an order of magnitude below the radio luminosity 
of the candidate TDE IGR J12580+0134 \citep[possibly due to an off-axis jet;][]{Irwin,Lei16}, and several 
orders of magnitude below the radio luminosities of relativistic TDEs like \swtd 
\citep[ascribed to an on-axis relativistic jet;][]{Zauderer11}. 
Additional radio observations of \srcname are planned and the time-evolution of the emission 
will be discussed in a future paper (Alexander et al. in prep).

\section{X-ray and UV variability}
\label{sec:var}

\begin{figure}
 \begin{center}
\begin{minipage}{5in}
    \rotatebox{0}{\includegraphics[width=9cm]{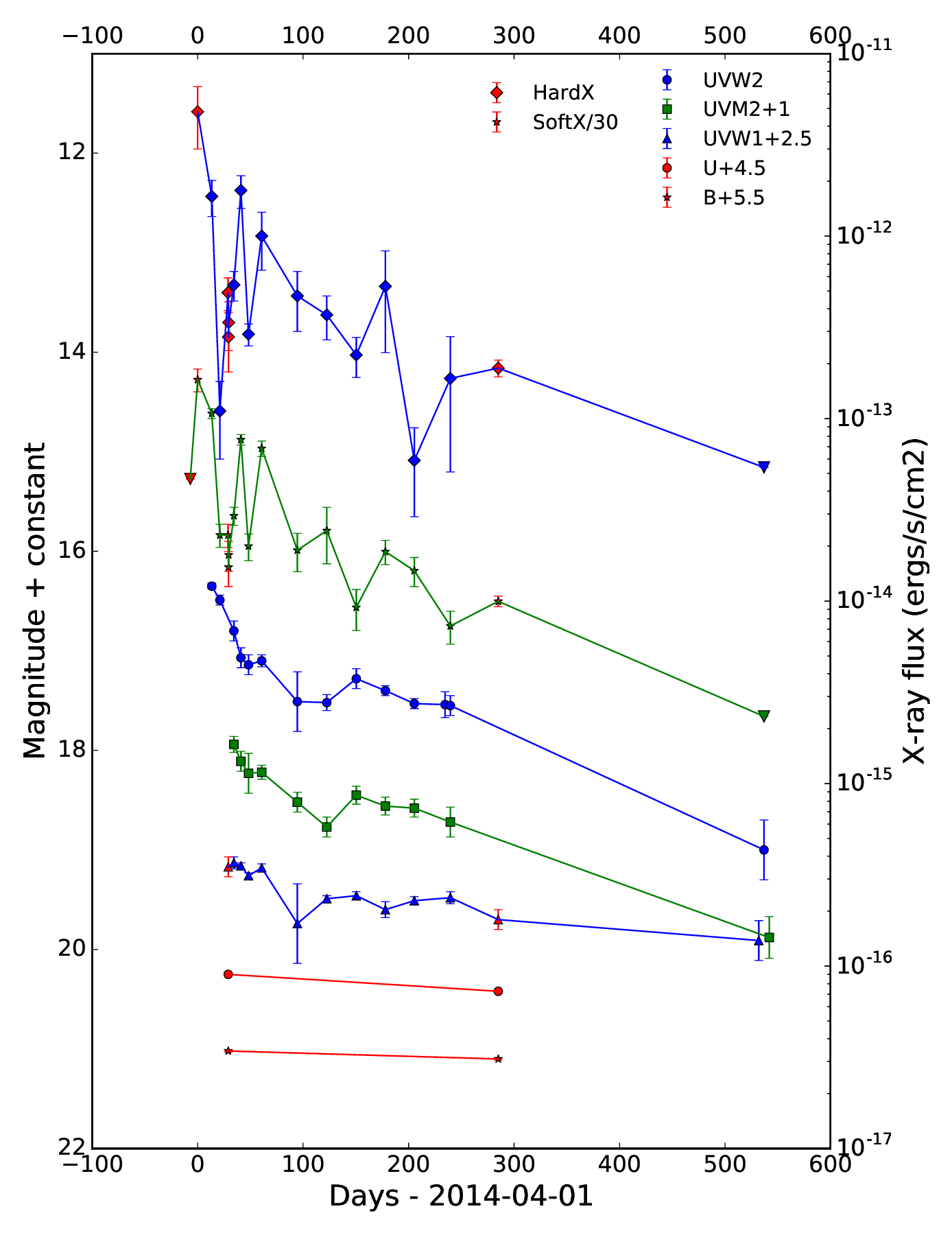}}
\end{minipage}

\begin{minipage}{5in}
    %\rotatebox{0}{\includegraphics[width=9cm]{lc_log_tight.eps}}
    \rotatebox{0}{\includegraphics[width=9cm]{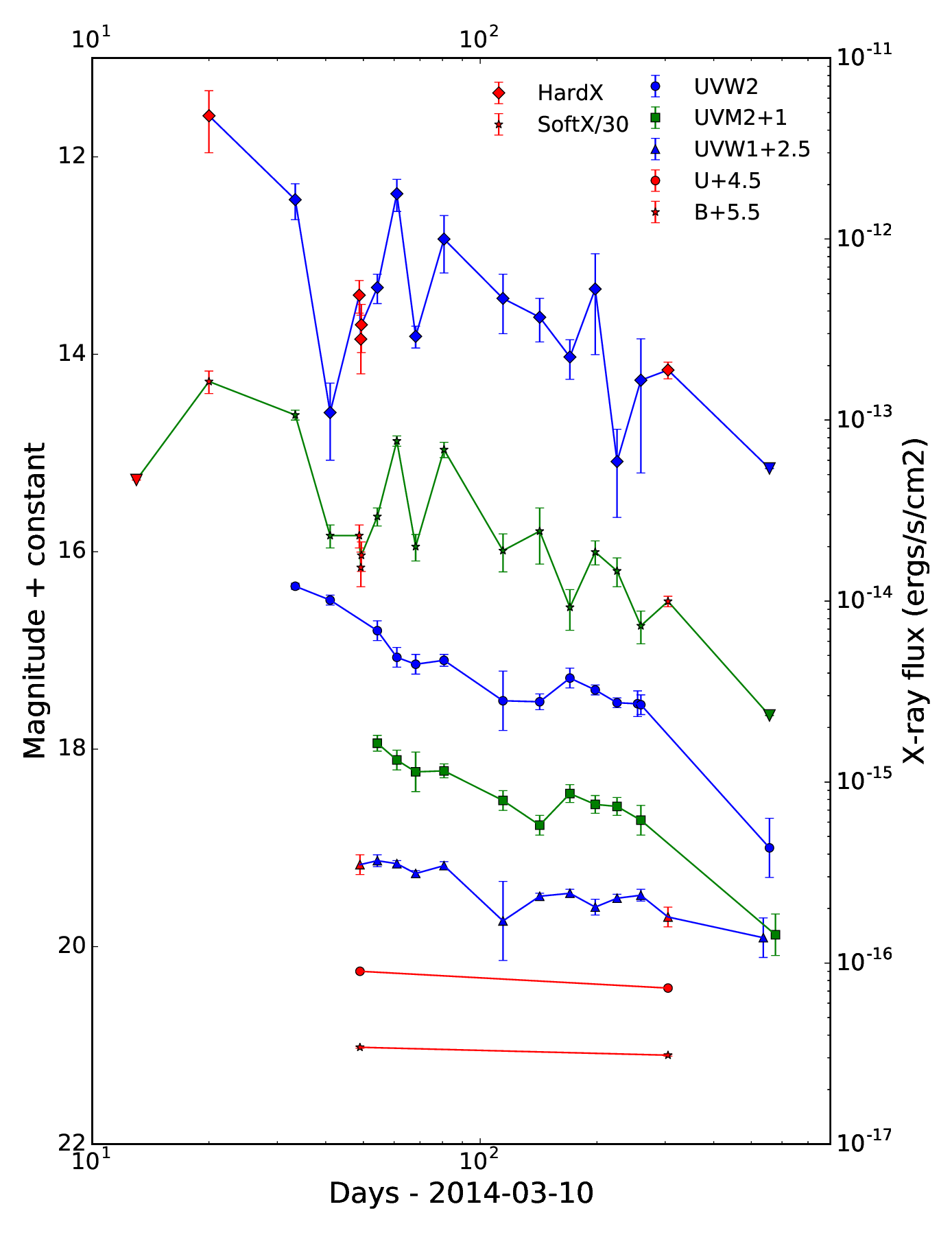}}
\end{minipage}

  \end{center}

\caption[\srcname longterm light curve]
{ \label{fig:lcurve} The long-term X-ray and UV light curve of \srcname
plotted against linear time (upper) and log time (lower) axes.
From top to bottom, 2-10 keV, 0.2-2 keV, Swift-uvw2 (1928\AA), Swift-uvm2 (2246\AA), 
Swift-uvw1/XMM-OM-UVW1 (2600\AA), XMM-OM-U (3480\AA), XMM-OM-B (4500\AA). \xmm points are marked in red and \swift points in blue or green. Upper limits are indicated by 
downwards triangles.
}
\end{figure}

In figure~\ref{fig:lcurve} we show the historical  
light curve of \srcnamens. All bands show 
gradually decreasing flux with the largest decrement being in the highest-energy bands.
%The indices of the flux decay with time, $f\propto t^{-x}$, are 
%$0.89\pm{0.39}$ (2-10 keV), $0.91\pm{0.28}$ (0.2-2 keV),
%$0.68\pm{0.14}$ (uvw2), $0.61\pm0.15$ (uvm2), $0.27\pm0.10$ (uvw1). 
%These decays are gentler than the index
%of 5/3 predicted by the classical model of TDE \citep{Rees88}. 
In principle, 
the date of peak emission can be constrained to be between an \xmm slew 
survey non-detection of 2014-03-24 and the detection on 2014-04-01 with 
a soft X-ray flux three times higher than this upper limit. 
Nevertheless, we see from this and other TDE \citep{Saxton12, Levan11} 
that the light curves can be very variable in the
early phase and so the actual peak may have occurred up to a 
few weeks earlier then the 2014-04-01 observed peak. 
The soft and hard X-ray light curves may be adequately hand-fit by 
the decay model predicted by the classical model of TDE 
\citep{Rees88, Phinney} of 
$L\propto(t-t_{0})^{5/3}$ with a peak time, 
$t_{0}=35$ days before the 
first detection (2014-02-25; see Fig.~\ref{fig:lcurvefit}).

\begin{figure}
  \begin{center}
\begin{minipage}{3in}
    \rotatebox{0}{\includegraphics[width=8cm]{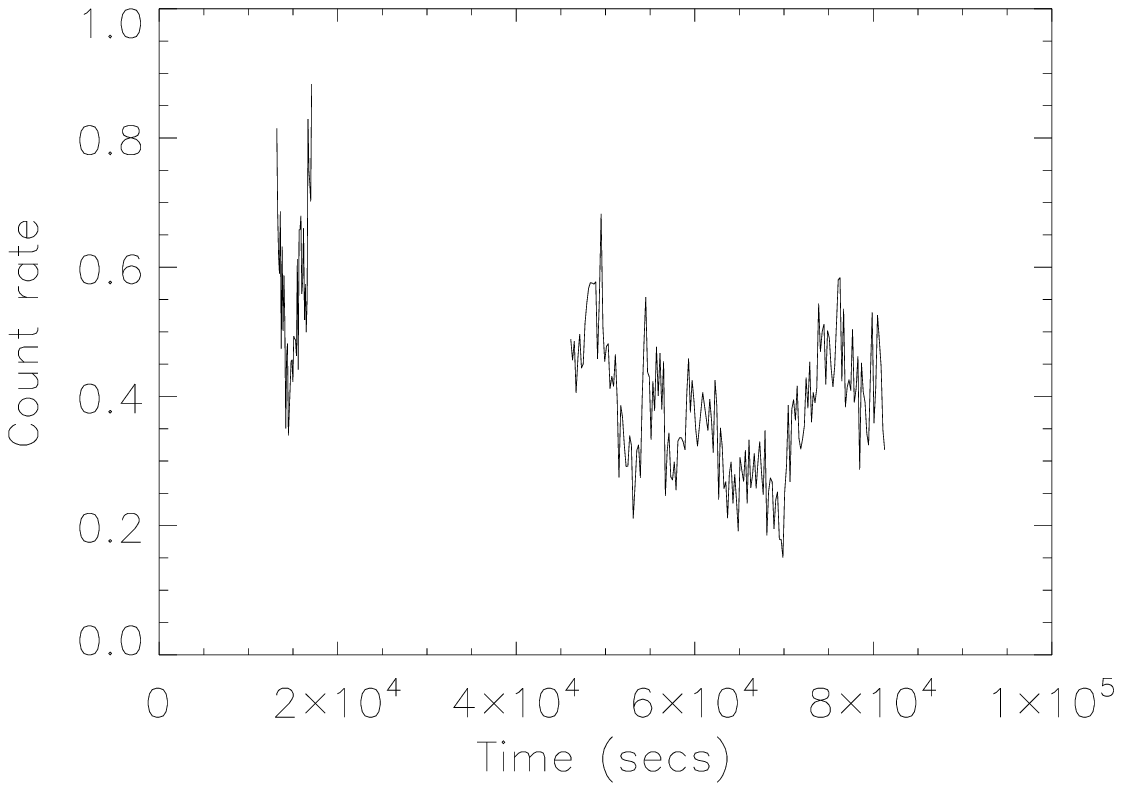}}
\end{minipage} 

\vspace*{0.3in}

\begin{minipage}{3in}
    \rotatebox{0}{\includegraphics[width=8cm]{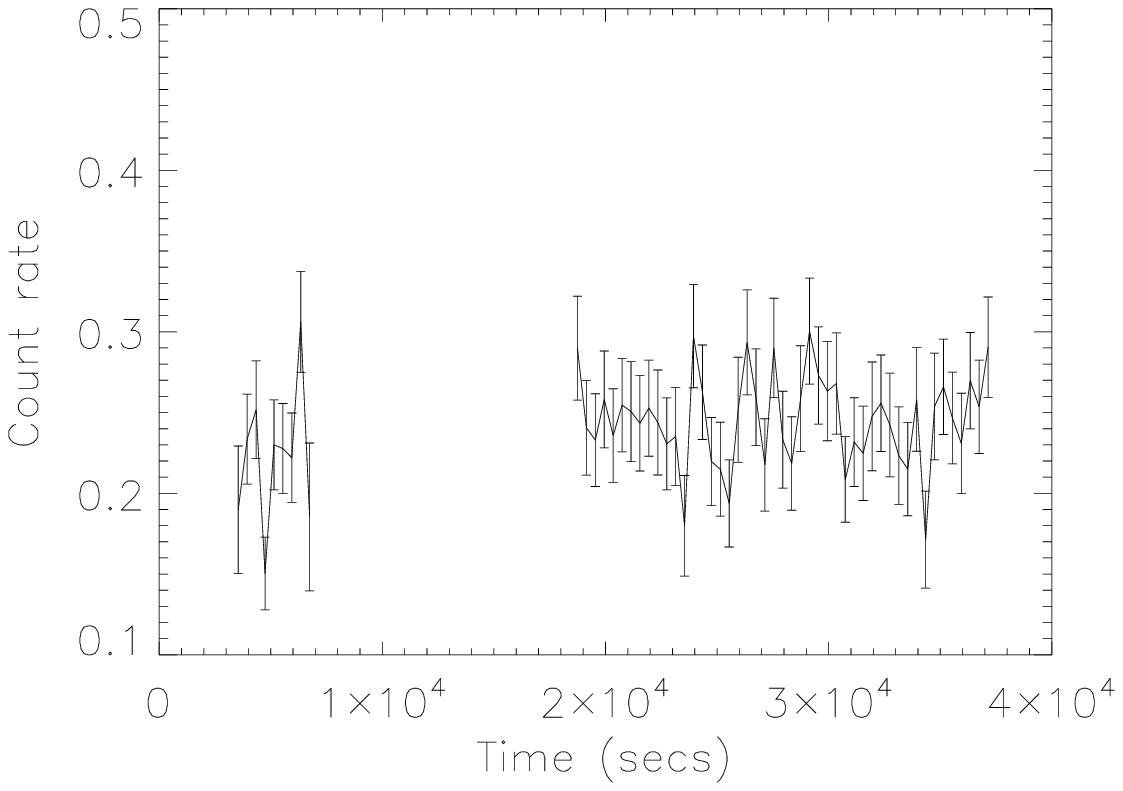}}
\end{minipage} 

  \end{center}
\caption[\srcname Short term light curve]
{ \label{fig:lcurve_short} The background subtracted, exposure corrected, 
EPIC-pn, 0.2--10 keV, light curve, for the 2014-04-30 (top) 
and 2015-01-12 (bottom) \xmm pointed observations.}
\end{figure}

\begin{figure}
\centering
\hspace*{-0.2in} \rotatebox{-90}{\includegraphics[height=7cm]{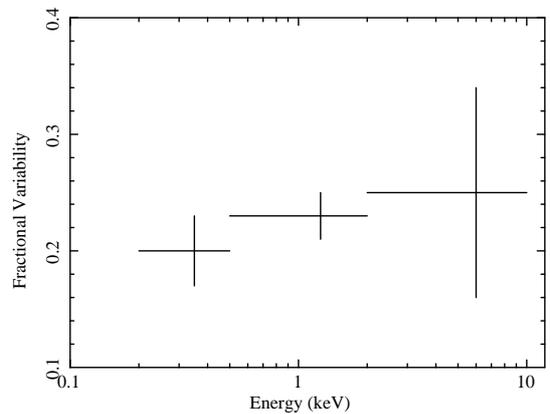}}
\caption[Fractional variability of \srcnamens.]
{ \label{fig:fracvar} The fractional variability of the EPIC-pn, 2014-04-30 data
of \srcname in the 0.2--0.5, 0.5--2 and 2--10 keV bands.}
\end{figure}

In Fig.~\ref{fig:lcurve_short}  we show the short-term light curve from the 
two \xmm pointed observations.
The X-ray flux, from the first XMM observation, shows large variations, with an amplitude up to a factor 4 on 
time scales of a few tens of ks. We exposure corrected the EPIC-pn light curve,
with the task {\em epiclccorr} and calculated the fractional variability in the
0.2--0.5, 0.5--2 and 2--10 keV net (background subtracted) light curves with {\em ekstest},
which uses algorithms developed in \citet{Edelson02} and \citet{Vaughan03}. The light curve shows 20--25\% variability, consistent within the errors, across the energy bands (Fig.~\ref{fig:fracvar}). 
Despite the apparent oscillations, a PSD analysis of the light curve
does not give a significant signal of periodicity. 

An estimate of the black hole mass may be obtained from the characteristics 
of the short-term variability. We have calculated the excess variance
of the variability, 
as defined in \citet{Ponti}, from 
three uninterrupted 10ks segments of the 2014-04-30 \xmm 
0.2--10 keV light curve, with 60 second binning. Using the scaling relations
of \citet{Ponti}, we find a black hole mass
of $M_{BH}=3.5^{+6.5}_{-2.4}\times10^{6}$\msolarns. 

The minimum doubling time of the flux in this observation is 
400 seconds, which
by light travel time arguments constrains the size of the emitting region 
to a radius of 400 light seconds or $1.2\times10^{13}$ cm. 
This is equivalent to $<$8--73 $R_{g}$ for a 
non-spinning black hole in this mass range.

The second observation is somewhat less variable; {\em ekstest} gives a fractional variability of 
$10.2\pm{3.0}$\% over the full 0.2--10 keV band. 
An excess variance analysis of a 10ks segment of the 0.2--10 keV light
curve of this observation, with 60 second binning, 
gives a weak constraint 
on the black hole mass of $M_{BH}>6\times10^{5}$\msolar (90\% confidence),
consistent with that obtained from the first observation.

% Fits to the long-term light curves
\begin{figure}
 \begin{center}
\begin{minipage}{5in}
    \rotatebox{90}{\includegraphics[width=7cm]{hardcurvefit.eps}}
\end{minipage}

\begin{minipage}{5in}
    %\rotatebox{0}{\includegraphics[width=9cm]{lc_log_tight.eps}}
    \rotatebox{90}{\includegraphics[width=7cm]{softcurvefit.eps}}
\end{minipage}

\begin{minipage}{5in}
    %\rotatebox{0}{\includegraphics[width=9cm]{lc_log_tight.eps}}
    \rotatebox{90}{\includegraphics[width=7cm]{uvcurvefit.eps}}
\end{minipage}

  \end{center}

\caption[\srcname longterm light curve models]
{ \label{fig:lcurvefit} Top: the hard X-ray light curve overlaid with a 
 model of $F_{2-10}=8\times10^{-14} \big[\frac{(\mathlarger{t} - 2014-02-25)}
{1 yr}\Big]^{-5/3}$\fluxUnitsns,
Middle: the soft X-ray light curve overlaid with a model of 
$F_{0.2-2}=1.1\times10^{-13} \big[\frac{(\mathlarger{t} - 2014-02-25)}
{1 yr}\Big]^{-5/3}$ \fluxUnitsns, Bottom: the 
Swift-uvw2 light curve, after subtraction of the galaxy background from
 a GALEX NUV observation of 2011-05-16, overlaid with a model of
$F_{UVW2}\propto\big[\frac{(\mathlarger{t} - 2014-02-25)}
{1 yr}\Big]^{-5/3}$(solid curve) and 
a better fitting model of $F_{UVW2}\propto\big[\frac{(\mathlarger{t} - 2014-02-25)}
{1 yr}\Big]^{-1.2}$ (dashed curve).
}
\end{figure}

%\Big[\big(\frac{R_{\rm{trunc}}}{R_{\rm{g}}}\big)^{5/4}

%Short-term light curves

\subsection{UV light curve}
\label{sec:uvlc}

During the two pointed \xmm observations, the optical monitor (OM)
cycled between the {\it B}, {\it U}, {\it UVW1} filters.
\swiftns-UVOT observations were
performed with the {\it uvw1, uvm2} and {\it uvw2} filters, except for the first two observations
which just used the {\it uvw2} filter.
The galaxy was detected in all of the filters.
Relative filter fluxes were determined using several nearby sources of
comparable brightness as references. The absolute flux scale was taken from 
the \swiftns-UVOT filters, with the \xmmns-OM UVW1 points scaled to these by 
a factor 1.5. 

In figure~\ref{fig:lcurve} we see that the flux, in all the UV filters, 
has responded to a strong flare coeval with the X-ray peak.
After the flare the flux declined in all the filters, with the uvw2
(2120\AA) reducing by a factor $12\pm{3}$ or 
$2.7\pm{0.3}$ magnitudes, to ${m=19.0}$ in 524 days.   

GALEX observed the galaxy on 2011-05-16 finding an NUV (2267 \AA)
magnitude of $20.14\pm{0.15}$ and
flux of $4.2\pm{0.6}\times10^{-13}$ ergs s$^{-1}$cm$^{-2}$.
The source was not detected in the FUV (1516 \AA) band.
The NUV filter has a similar bandpass to the uvw2 filter and
its flux is one magnitude lower than the last uvw2 observation of \swiftns,
suggesting that the flux was still decaying in this filter on 2015-09-25.
We can use the GALEX measurement to subtract the contribution of 
the galaxy 
from the {\it uvw2} observations of the flare. The resulting {\it uvw2}
 light curve (shown in Fig.~\ref{fig:lcurvefit}) has a shallower decay
index than the canonical $t^{-5/3}$ and may be better modelled with an index 
$t^{-1.2}$ (Fig.~\ref{fig:lcurvefit}).

%The magnitude of the flare decreases with increasing wavelength, with the
%U filter decreasing by $\Delta U=0.17\pm{0.02}$ and B filter by 
%$\Delta B=0.08\pm{0.02}$ magnitudes between the first and second XMM pointed observations.

The {\it uvw2, uvm2} and {\it uvw1} light curves show signs of a 
recovery in flux around day 180, which may be echoed in the X-ray curves. 
Similar secondary bumps have been noted in candidate
TDE light curves before \citep[e.g.][]{Merloni}.

%\begin{figure}
%\centering
%\rotatebox{-90}{\includegraphics[height=9cm]{optfilt.eps}}
%\hspace*{-0.2in} \rotatebox{0}{\includegraphics[height=7cm]{uvcurve.eps}}
%\caption[UV light curve of \srcnamens.]
%{ \label{fig:uvlcurve} The UV light curve of \srcname
%taken from the \swiftns-UVOT and \xmmns-OM observations using the filters:
%{\it U} (3480\AA), {\it UVW1} (2950\AA), {\it UVW2}
%(2120\AA) and {\it UVM2} (2340\AA)}
%\end{figure}

\subsection{X-ray spectral analysis}
\label{sec:specfit}
We extracted spectra from the first \xmm pointed observation (hereafter XMM1),
taken 29 days after the peak flux seen on 2014-04-01, from circles with
optimum radii determined by the task {\em eregionanalyse}
for the EPIC-pn and MOS-1 cameras\footnote{MOS-2 data were not used in 
this analysis as the small window mode data have lower statistics than MOS-1.}.
Background spectra were created from nearby source-free regions\footnote{ 
As the observation was split into three segments 
(see section~\ref{sec:xuvobs}), source and
background spectra and effective area files were
extracted individually for each segment and combined using the ftools tasks
{\em mathpha} and {\em addarf}.}.
The source spectra were grouped to have a minimum of 25 counts
per bin and were fitted simultaneously within the XSPEC
package (v12.8.2). 
Fits were performed using the Chi-squared statistic over the energy range  
0.3--10 keV, using a constant to account for
the small differences in normalisation between the instruments.
The XMM1 observation had 16500 background 
subtracted counts in the two cameras and in this energy range, while 
the second pointed observation 
(hereafter XMM2) had 4700 background subtracted counts. Quoted errors
are 90\% confidence unless otherwise stated.
 
As a first step we fit the XMM1 observation with a simple power-law model 
and galactic absorption of $1.5\times10^{21}$cm$^{-2}$, modelled by TBABS with element abundances
set to those in \citet{Wilms}.
The fit is poor (\chired=1774/399) with a slope of $\Gamma\sim4$.
If we fit over the 2--10 keV range only, the slope flattens to $1.76\pm{0.11}$
and large residuals are present at lower energies).

We attempted to model the low-energy excess as an extra emission component,
most successfully with a blackbody component of $kT=86\pm{2}$eV (\chired=415/397).
Neither intrinsic absorption (\chired=460/396) nor relativistically-blurred reflection from an ionised disc (\chired=664/396) gave as good a fit  (see table~\ref{tab:specfits} for details). Good fits could be obtained by adding a further component. A power-law
with two intrinsic absorbers (\chired=393/393) or a power-law plus a black-body 
with an intrinsic absorber (\chired=383/394) modelled the 
X-ray spectrum adequately.   

A black-body temperature of kT=86 eV is significantly lower than the $kT\sim120$ eV equivalent
temperature ubiquitously found for the soft-excess component of AGN \citep[e.g.][]{Done12} but is similar to
the temperature of thermal emission which might be expected from material in the inner 
disk about a black hole of $10^{6}$\msolar. Based on this we included the UV filters, 
OM-UVW1 and \swiftns-uvm2 in the fit, with intrinsic absorption given by $B-V=0.14$ \citep{schlaf}, to test whether they
represent the low-energy part of disc emission. The \swiftns-uvm2 filter 
flux was taken from a linear extrapolation of the uvm2 light curve to 
the date of the XMM1 observation (the first uvm2 measurement was 
actually taken 5 days later). As these two UV filters
show large flares, they are clearly dominated by nuclear emission, and no attempt was made to subtract any emission from the stellar population. 

\begin{figure}
\centering
%\rotatebox{-90}{\includegraphics[height=9cm]{resplot_ob1_pn.eps}}
%\rotatebox{-90}{\includegraphics[height=9cm]{resplot_pow.eps}}
\rotatebox{-90}{\includegraphics[height=9cm]{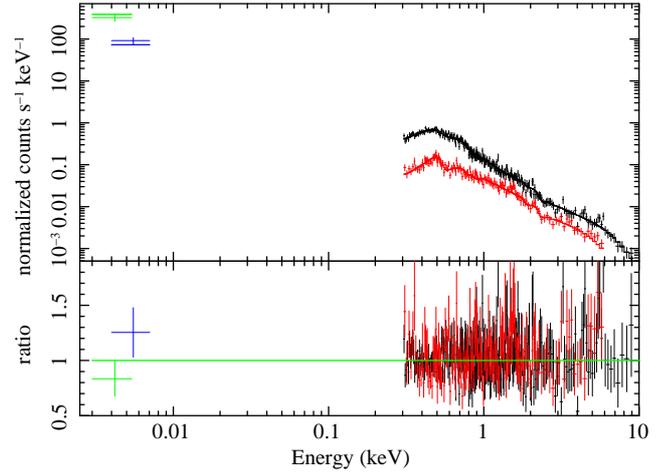}}
\caption[A fit of a multi-colour disc model (optxagn) to the first \xmm observation]
{ \label{fig:optxagnfit} A fit of a multi-colour disc model (optxagn), 
absorbed by the Galactic column, to the EPIC-pn (black),  
MOS-1 (red) and OM-UVW1 (blue) data from the \xmm pointed
observation of 2014-04-30 and to the Swift-uvm2 filter flux, extrapolated to 
the same date (green; see text).}
\end{figure}

We changed to the  {\tt optxagnf} multi-coloured disc model \citep{Done12}
which assumes that the disc emits a colour-corrected blackbody down to a given 
radius, within which a two-phase plasma forms a corona. The model is
energetically self-consistent with accretion energy powering thermal emission from 
the disc, a soft X-ray excess produced by Comptonisation in optically thick plasma, and a high-energy power law produced by Comptonisation in an optically thin plasma.
We  first assumed  a  pure  thermal  accretion  disc model,
with a power-law component but no soft excess from optically-thick Comptonisation but
were unable to get a good fit.  
We then allowed the optically thick Comptonisation to vary and found an acceptable
fit with the soft X-ray excess, over a hard power-law, being provided by both thermal
emission and the Comptonisation 
component  (\chired=404/395; Fig.~\ref{fig:optxagnfit}; Table~\ref{tab:optxagn}). In this fit the black hole mass,  
is only very loosely constrained to $M_{BH}<2\times10^{7}$\msolar (90\% confidence) and so 
we fixed it to the best fit value from the variability 
analysis, $M_{BH}=3.5\times10^{6}$\msolarns. A non-spinning black hole was assumed ($a=0$). 
Although the fit is quite good, an excess is clearly present in the residuals between 4--6 keV. 
We attempted to model this with a distant reflector, using the {\em pexmon} \citep{Nandra07} model. 
This gave a small improvement of 
 $\delta\chi^{2}=5$ for two degrees of freedom (dof), with the redshift frozen to that of the galaxy.
As the residuals appear to be broad we then tried the diskline model  \citep[Laor; ][]{Laor91} which gave a
better fit ($\delta\chi^{2}=22$ for two dof; F-test=99.998\%) 
with a line energy of 5.85 keV in the source 
rest frame, which would indicate a redshift of $\sim0.09$ if the line is due to neutral
Fe K$_{\alpha}$. As an alternative we fit the spectrum with a 
partially-covering 
intrinsic ionised absorber 
\citep[zxipcf;][]{reeves} which also gave a good result ($\delta\chi^{2}=19$ for three dof; F-test=99.989\%). 
%{\bf An F-test gives a confidence of 99.998\% and 99.989\% for the inclusion of the extra model component respectively}.
%The black hole mass from this fit is $M_{BH}=3^{+14}_{-1.5}\times10^{6}$\msolar (90\% confidence)
For this model, the accretion rate is $\dot{m}=0.047^{+0.009}_{-0.011}$ $\dot{m}_{edd}$ for a Schwarzchild black hole of $M_{BH}=3.5\times10^{6}$\msolarns. 
%or $\dot{m}=0.021^{+xxx}_{-yyy} \dot{m}_{edd}$  for a Kerr (a=0.99) maximally-spinning BH. 
Note that the accretion rate during the 2014-04-01 slew observation will have been roughly 10 times higher than this.

\begin{figure}
\centering
\rotatebox{-90}{\includegraphics[height=9cm]{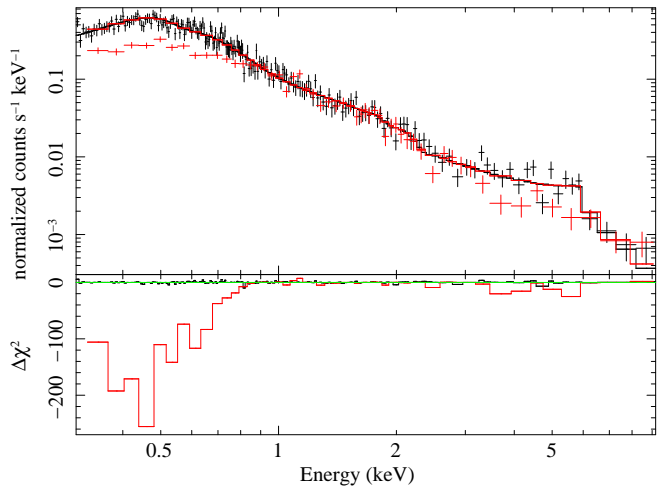}}
\caption[A comparison of spectra from the two \xmm pointed observations.]
{ \label{fig:doublespec} A comparison of EPIC-pn spectra from the XMM1 and XMM2 pointed 
observations of 2014-04-30 (black) and 2015-01-12 (red). The lower panel shows
the residuals from the model fit, TBABS*optxagn*zxipcf to the 
2014-04-30 spectrum.}
\end{figure}

If we now look at the XMM2 spectra, taken 285 days after discovery, then we can see that the major change is a reduction in the observed flux below 2 keV
and a reduction, or removal, of the 4--6 keV excess seen in XMM1   (Fig.~\ref{fig:doublespec}). We again simultaneously fit the EPIC-pn, MOS-1, 
OM-UVW1 and extrapolated \swiftns-uvm2 datasets with the {\em optxagn} model and the same
fixed values of $M_{BH}$ and spin.
The fit is reasonable, \chired=117/105. In this case no significant 
improvement was found
by adding an additional intrinsic absorption or broad line component 
(Tab.~\ref{tab:optxagn}). From the fits we find that the accretion rate has 
reduced in XMM2 to $\dot{m}=0.015^{+0.008}_{-0.006}$ $\dot{m}_{edd}$. 
% Rcor can be conatrained to >10 in all fits.

%The temperature remained consistent between the observations at xxxx K.

\begin{figure}
  \begin{center}
    \rotatebox{0}{\includegraphics[width=9cm]{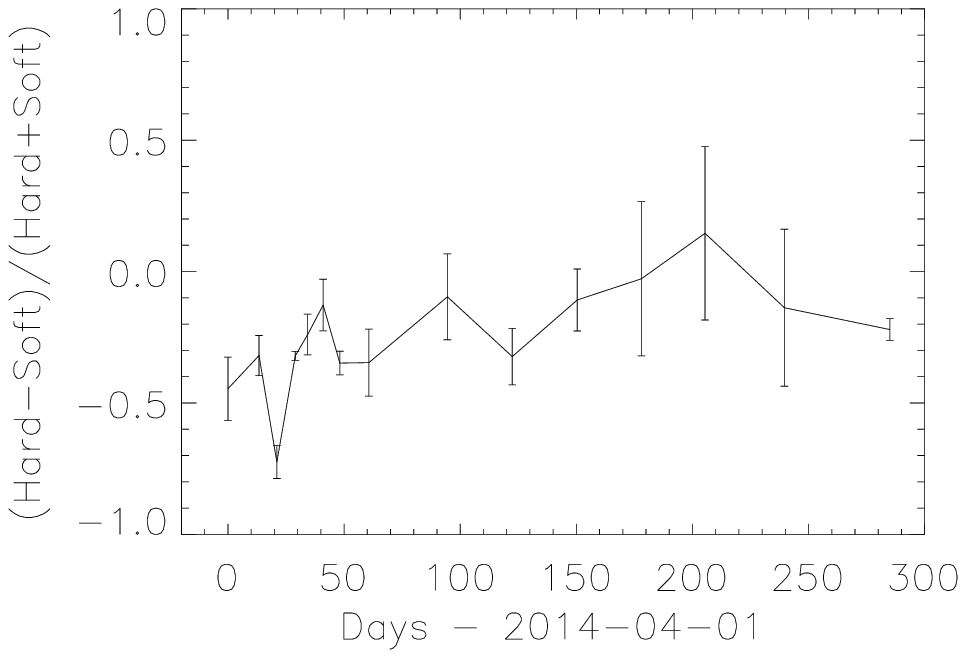}}
  \end{center}
\caption[\srcname X-ray hardness ratio]
{ \label{fig:hratio} The 2--10 keV to 0.2--2.0 keV flux ratio, 
calculated by fitting a phenomenological model of a
 power-law plus  black-body absorbed by
the Galactic column.}
\end{figure}

\renewcommand{\arraystretch}{1.5}
\begin{table*}[ht]
{\small
\caption{Spectral fits to \xmm pointed observations of \srcnamens.}
\label{tab:specfits}      % is used to refer this table in the text
\begin{center}
\begin{small}
\begin{tabular}{c c| c | c | c | cc| ccc |c}
\hline\hline                 % inserts double horizontal lines
% \multicolumn{7}{c}{emission model} & \multicolumn{3}{c}{intrinsic absorption} & C/dof  \\
\\
%\multicolumn{2}{c}{Power-law$^{a}$}  & Distant Reflection$^{b}$ & \multicolumn{3}{c}{zxipcf~$^{c}$} & \multicolumn{2}{c}{Ionized Reflection$^{d}$} & C/dof~$^{e}$ \\
%$\Gamma$ & Norm  &  Norm & $N_{H}$ & $xi$ & cf & $xi$ & Norm & \\
 %        & keV$^{-1}$ cm$^{-2}$ s$^{-1}$ & keV$^{-1}$ cm$^{-2}$ s$^{-1}$ & 10$^{22}$ cm$^{-2}$ & log & \% & log & cm$^{-2}$ s$^{-1}$ & \\
\multicolumn{2}{c}{Power-law$^{a}$}  & Power-law$^{b}$ & 
Bbody$^{c}$ & Brem$^{d}$ &   \multicolumn{2}{c}{Reflection$^{e}$} & \multicolumn{3}{c}{zxipcf~$^{f}$} & C/dof~$^{g}$ \\
$\Gamma$ & Norm  &  $\Gamma$ & kT & kT & Norm & $xi$ & $N_{H}$ & $xi$ &  cf & \\
         & keV$^{-1}$ cm$^{-2}$ s$^{-1}$ & & eV & eV & $\times10^{-8}$ & erg cm/s & 10$^{22}$ cm$^{-2}$ & log & \% & \\
\hline\noalign{\smallskip}
\multicolumn{9}{c}{\xmm observation 1 - 2014-04-30 (XMM1)} \\
\hline
% Power-law + pow fit
$4.61\pm{0.08}$ & $1.26\pm{+0.08}\times10^{-4}$ &    $1.43\pm{0.09}$ & 
&   &     &    &    &   &  &  516/397  \\
% Power-law + bbod fit
$1.94\pm{0.06}$ & $1.13\pm{0.05}\times10^{-4}$ &   & $86\pm{2}$ & 
&   &     &    &    &   &  415/397  \\
% Power-law + brem fit
$1.82\pm{0.06}$ & $0.98\pm{0.06}\times10^{-4}$ &   & & $170\pm{5}$ & 
&   &     &    &    &   416/397  \\
% Power-law + reflionx*kdblur fit
$1.48\pm{0.06}$ & $5.8\pm{0.4}\times10^{-5}$ & &  &   & $2.4^{+1.4}_{-0.4}$ 
& $855^{+150}_{-260}$ &     &   &   & 664/396  \\
% Power-law + zxipcf
$3.31\pm{0.04}$ & $1.72\pm{0.21}\times10^{-3}$ &   & &  & & & $17.5^{+0.9}_{-1.0}$ & $1.95\pm{0.02}$ & $92\pm{1}$  &  460/396  \\
% Power-law + bbod * zxipcf fit 
$2.45\pm{0.20}$ & $4.0^{+1.9}_{-1.3}\times10^{-4}$ & & $86^{+4}_{-3}$ &     &     &  &  $18.0^{+4.2}_{-2.9}$ & $1.96^{+0.08}_{-0.38}$  &  $68^{+10}_{-13}$  & 
383/394  \\ 
% Power-law + pexmon + reflionx*kdblur
%$1.92\pm{0.02}$   & $4.46^{+0.07}_{-0.08}\times10^{-4}$  & %$2.48^{+1.66}_{-1.55}\times10^{-4}$  &    &   &   &   $10.0^{+2.7}_{-10.0}$ &  %$1.03^{+0.18}_{-0.27}\times10^{-6}$ & 3471/3227  \\
\noalign{\smallskip}\hline
\multicolumn{9}{c}{\xmm observation 2 - 2015-01-12 (XMM2) } \\
\hline\noalign{\smallskip}
% Power-law + pow fit
$4.15^{+0.42}_{-0.33}$ & $6.7^{+2.8}_{-2.1}\times10^{-4}$ &  $1.97\pm{0.21}$ & 
&   &     &    &    &   &  &  115/105  \\
% Power-law + bbod fit
$2.38\pm{0.11}$ & $1.10^{+0.11}_{-0.09}\times10^{-4}$ &   & $85\pm{7}$ & 
&   &     &    &    &   &  122/105  \\
% Power-law + brem fit
$2.28\pm{0.13}$ & $1.04^{+0.12}_{-0.12}\times10^{-4}$ &   & & $177\pm{24}$ &
&   &     &    &    &    116/105  \\
% Power-law + xillver fit
%$1.98\pm{0.02}$ & $4.83^{+0.05}_{-0.04}\times10^{-4}$ & &  &   & $16^{+23}_{-13}$ & $1.67^{+0.77}_{-0.72}$ &     &   &   & 705/397  \\
% Power-law + zxipcf
$3.07\pm{0.06}$ & $5.2^{+2.1}_{-1.5}\times10^{-4}$ &   & &  & & & $14^{+4}_{-9}$ & $1.8^{+0.8}_{-1.4}$ & $74^{+7}_{-11}$  &  122/104  \\
% Power-law + bbod * zxipcf fit
$2.79\pm{0.24}$ & $2.84^{+1.07}_{-1.04}\times10^{-4}$ & & $76^{+11}_{-19}$ &     &     &  &  $6^{+108}_{-4}$ & $-0.55^{+2.38}_{-0}$  &  $61^{+18}_{-38}$  & 116/102  \\
\noalign{\smallskip}\hline
\end{tabular}
\\
\end{small}
\end{center}
All fits include absorption by the Galactic column (model {\tt TBABS}, $N_{H}=1.5\times10^{21}$cm$^{-2}$, abundances from \citet{Wilms}). Errors are 90\% confidence. \\
$^{a}$ Continuum power-law with photon index and normalisation, $^{b}$ Slope of secondary power-law modelling high-energy emission, $^{c}$ black-body model temperature, $^{d}$ Bremsstrahlung model temperature, $^{e}$ ionized reflection \citep[{\tt reflionx};][]{RossFabian} 
with ionization state and normalisation, convolved with a relativistic blurring
model \citep[{\tt kdblur};][]{Laor91,Fabian2002}, with fixed emissivity index=3, inner radius=4.5 $GM/c^{2}$, outer radius=100 $GM/c^{2}$ and inclination=$45\,^{\circ}$,  
 $^{f}$ ionized absorption \citep[{\tt zxipcf};][]{reeves} with equivalent column density, ionization state and covering fraction,  
 $^{g}$ C-statistic / number of degrees of freedom. 
}
\end{table*}

\begin{table*}[h]
{\small
\caption{Fits with a multi-colored disk model to \xmm pointed observations of \srcnamens.}
\label{tab:optxagn}      % is used to refer this table in the text
\begin{center}
\begin{small}
\begin{tabular}{c c c c c|cc| ccc|c}
\hline\hline                 % inserts double horizontal lines
% \multicolumn{7}{c}{emission model} & \multicolumn{3}{c}{intrinsic absorption} & C/dof  \\
\\
%\multicolumn{2}{c}{Power-law$^{a}$}  & Distant Reflection$^{b}$ & \multicolumn{3}{c}{zxipcf~$^{c}$} & \multicolumn{2}{c}{Ionized Reflection$^{d}$} & C/dof~$^{e}$ \\
%$\Gamma$ & Norm  &  Norm & $N_{H}$ & $xi$ & cf & $xi$ & Norm & \\
 %        & keV$^{-1}$ cm$^{-2}$ s$^{-1}$ & keV$^{-1}$ cm$^{-2}$ s$^{-1}$ & 10$^{22}$ cm$^{-2}$ & log & \% & log & cm$^{-2}$ s$^{-1}$ & \\
\multicolumn{5}{c}{optxagnf$^{a}$}  & \multicolumn{2}{c}{Disk line$^{b}$} & \multicolumn{3}{c}{zxipcf~$^{c}$} & C/dof \\
\hline
 $\dot{m}$   & slope & kT & $\tau_{e}$ & $f_{pl}$ & $E_{line}$ & Norm & $N_{H}$ & $xi$ &  cf & \\
 log($\dot{m}_{edd}$) & & (eV) & & & (keV) & $\times10^{-6}$ & 10$^{22}$ cm$^{-2}$ & log & \% & \\
\hline\noalign{\smallskip}
\multicolumn{10}{c}{\xmm observation 1 - 2014-04-30 (XMM1)} \\
\hline
% optxagnf
$-1.37^{+0.12}_{-0.16}$   & $1.87\pm{0.07}$ & $101^{+9}_{-8}$ & $33^{+19}_{-8}$ & $0.25^{+0.09}_{-0.10}$ & & & & & & 404/395 \\

%optxagnf + laor
 $-1.36^{+0.12}_{-0.17}$  & $2.06\pm{0.10}$ & $92^{+9}_{-5}$ & $49^{+51}_{-20}$ & $0.28^{+0.09}_{-0.03}$ & $5.76^{+0.17}_{-0.14}$ & $5.36^{+1.81}_{-1.86}$ & & & & 382/393 \\

%optxagnf + zxipcf
 $-1.33^{+0.08}_{-0.11}$  & $1.84\pm{0.09}$  & $153^{+27}_{-28}$ & $17^{+5}_{-3}$ & $0.13^{+0.05}_{-0.07}$ & & &
$0.7^{+1.8}_{-0.3}$ & $2.16^{+0.21}_{-0.33}$ & $>63$ & 383/392 \\
%$0.9^{+3.1}_{-0.3}$ & $2.17^{+0.15}_{-0.14}$ & $100^{+0}_{-40}$ & 395/392 \\

\noalign{\smallskip}\hline
\multicolumn{11}{c}{\xmm observation 2 - 2015-01-12 (XMM2)} \\
\hline\noalign{\smallskip}
% optxagnf
 $-1.81^{+0.18}_{-0.24}$   & $2.30\pm{0.12}$ & $114\pm{20}$ & $24^{+19}_{-6}$ & $0.24^{+0.24}_{-0.12}$ & & & & & & 117/105 \\

%optxagnf + laor
 $-1.87^{+0.16}_{-0.24}$  & $2.37\pm{0.13}$ & $106^{+6}_{-4}$ & $25^{+52}_{-7}$ & $0.28^{+0.42}_{-0.13}$ & $6.84^{+0.4}_{-0.3}$ &  $1.7^{+2.0}_{-1.7}$& & & & 116/103 \\

%optxagnf + zxipcf
 $-1.83^{+0.2}_{-0.2}$  & $2.32^{+0.20}_{-0.17}$ & $110^{+35}_{-15}$ & $25^{+25}_{-8}$ &  $0.27^{+0.28}_{-0.19}$ & & &
$15^{+10}_{-8}$ & $<6$ & $<1$ & 117/102 \\
%$15^{+10}_{-8}$ & 2.00 & $0^{+1}_{-0}$ & 117/102 \\

\noalign{\smallskip}\hline
\end{tabular}
\\
\end{small}
\end{center}
All fits include absorption by the Galactic column (model {\tt TBABS}, $N_{H}=1.5\times10^{21}$cm$^{-2}$). Errors are 90\% confidence. \\
$^{a}$ Shown are parameters of the optxagnf model \citep{Done12} with black 
hole mass fixed to $M_{BH}=3.5\times10^{6}$\msolar and spin to a=0; 
accretion rate, power-law slope, electron temperature of the soft 
Comptonisation component, optical depth, the disc radius, within which emission
changes from black-body to a Comptonised spectrum, and the fraction of energy emitted
as a power-law. $^{b}$ observed energy of
a relativistically broadened line given by the \citep[Laor model; ][]{Laor91}. 
$^{c}$ ionized absorption with equivalent column density, ionization state 
and covering fraction ({\em zxipcf}).
}
\end{table*}

\renewcommand{\arraystretch}{1.0}

\section{Discussion}

The lack of emission lines in the optical spectrum taken during the 2MASS redshift
survey make it unlikely that \srcname was a steadily-emitting AGN before the
2014 flare. We have confirmed this diagnosis with a further optical spectrum which 
allows us to constrain the luminosity of the [OIII]$\lambda$5007 line to 
$L_{OIII}<4\times10^{-15}$\fluxUnitsns. The unabsorbed 2--10 keV flux from 
the \xmm slew observation of 2014-04-01, $F_{2-10 keV}\sim4.8\times10^{-12}$\fluxUnitsns, 
gives a ratio $L_{2-10 keV}$ / $L_{OIII}>1000$, well in excess of the ratio $\sim 1 - 100$
found in samples of Seyfert galaxies \citep{Panessa06,Lamastra09}.
The WISE colours, W1 - W2 = 0.0, are also suggestive of a non-active
galaxy \citep{Stern12}.

With only a single epoch of observations, an association between the radio emission and \srcname
 is necessarily tentative. There are no archival radio observations of the host of \srcname but from the above analysis it is unlikely that the radio emission is due to AGN activity. The stellar population synthesis modeling discussed in 
section~\ref{sec:optobs} suggests that the radio emission is also 
unlikely to be due to ongoing star formation. The upper limit on OIII 
emission of $4\times10^{-15}$\fluxUnits corresponds to a star formation 
rate of $< 0.4^{+0.6}_{-0.3}$ \msolar/year \citep{Moustakas}. 
This rate implies a radio flux density of 
 $< 0.5^{+0.7}_{-0.3}$ mJy at 1.5 GHz \citep{Condon}. A more restrictive 
estimate of the current star formation rate can be obtained from 
the 2011 NUV detection of the host galaxy by GALEX (section~\ref{sec:uvlc}). 
Following the method of \citet{Wilkins12}, we estimate the star formation rate to be $\sim 0.02$ \msolar/year. This rate implies a 1.5 GHz radio 
flux density of $\sim 0.03 \pm 0.01$ mJy, which is a factor of 40 less than the 
flux we observe. 

Scaling the best fitting spectral model from the XMM1 observation to the 
2014-04-01 \xmm slew count rate gives $L_{bol}\sim2\times10^{44}$\lumUnitsns.
As this is in excess of
even the brightest ULX, the conclusion is that the SMBH of \srcname has experienced a sudden increase in available fuel, probably from the tidal disruption of a stellar
object. 

From the relationship of black hole mass to bulge K-band luminosity
\citep{MarconiHunt} we find $M_{BH}\sim6\times10^{6}M_{\odot}$ using the point source
2MASS magnitude of $m_{K}=11.72$, with a systematic error of 0.3 dex 
($2\times10^{6}M_{\odot}$).
%From the optical absorption lines width we find a value of 
%{\boldmath $M_{BH}\sim3\times10^{7}M_{\odot}$} 
This is consistent with the value of  
$M_{BH}=3.5^{+6.5}_{-2.4}\times10^{6}$\msolar derived from
the X-ray variability study (section~\ref{sec:var}).
%while the spectral fitting gave a good fit with $M_{BH}=1.5--17\times10^{6}$\msolar(section~\ref{sec:specfit}). 
%We take $M_{BH}\sim4\times10^{6}$\msolar as a working measure. 

The total energy release, integrating between the discovery date and the XMM2 
observation 285 days later is $\sim5\times10^{50}$ ergs, equivalent to a
consumed mass of 0.025\msolar assuming a conversion efficiency of 0.1.

%The inconsistency of the UV and X-ray emission 
The fact that we cannot fit the UV and soft X-ray excess with a single thermal disk model
implies that they are not part of the same 
thermal component. This is in common with many TDE where the UV emission appears to be the high-energy part of a separate, lower-temperature, optical component \citep{Gezari12,Holoien16a,Holoien16b}. 
%Only in the candidate TDE 2MASS 0619-65 has this been seen before \citep[are there others...][]{Saxton14}. 

In the late stages of a flare the X-ray spectrum has been seen to harden on
several occasions \citep{Vaughan04,KomossaBade,Komossa1242}. 
At peak, however, most TDE discovered in the soft X-ray band did
not have significant emission above $\sim2$ keV 
\citep[e.g.][]{KomossaBade,Esquej08,Saxton12,Lin15}.
\srcname has comparable flux in the soft and hard X-ray bands 
(Fig.~\ref{fig:hratio}) and is a clear example, without significant 
internal absorption, of a TDE with strong thermal {\it and} power-law emission. 
%\srcname has many properties in common with the well-monitored TDE, ASSASN 14li (\refs); variable UV, soft X-ray and radio emission. They differ in that ASSASN 14li
%showed a strong optical flare, with no hard X-ray emission, while \srcname exhibited
%a very mild change in B flux but a strong power-law. etc etc

It is not yet completely demonstrated whether power-law X-ray emission in
TDE is produced in a jet or an AGN-like comptonisation region. While the
luminosity of \srcname is below that of the proposed \swift on-axis
relativistic jets it is comparable with NGC 4845, whose hard  
\citep[$\Gamma$$\sim$2.2;][]{Walter13} 
X-ray emission has been ascribed to an off-axis jet \citep{Lei16} based
on its radio emission \citep{Irwin}. The detection of radio emission
from \srcname may point in this direction. However, the detection of
spectral features around 6 keV, whether due to intrinsic absorption 
or ionised reflection, are more suggestive of non-jetted intrinsic
emission. In this context the radio emission could be interpreted as a
non-relativistic outflow akin to that seen in \assasn \citep{Alexander16}.
The nature of the radio emission in \srcname will be explored further 
in Alexander et al. (in prep.). 

It has recently been discovered that a large fraction of 
optically-selected TDE occur in host galaxies which 
have experienced a starburst 
that has recently finished \citep{Arcavi14,French}. 
From detailed stellar population modelling in section~\ref{sec:optobs} 
we showed that the host galaxy of
\srcname has undergone a period of strong star formation
about 2 Gyr ago. \citet{French} use line diagnostics to identify these
galaxies by the presence of an
$H_{\delta}$ absorption line with equivalent width (EW) $>3\AA$ 
and an $H_{\alpha}$ emission line with $EW<3\AA$. 
We measure an EW, in absorption, of $H_{\delta}\sim3\AA$ from the 
CTIO-2MASS (pre-flare) spectrum and determine a strict 
upper limit to the $H_{\alpha}$ emission of $<1\AA$, 
from the, higher statistic, Las Campanas spectrum after correcting for
stellar absorption. Therefore, this TDE also appears 
to be hosted in a rare post-starburst galaxy.

\section{Summary}

A flare was detected from the galaxy \srcname on April 1st 2014,
reaching a bolometric luminosity, $L_{bol}=2\times10^{44}$ \lumUnitsns.
The source flux subsequently decayed by a factor $>70$
in the 2--10 and 0.2--2 keV X-ray bands and by up to 2.7 magnitudes in 
UV filters. The flux change in the optical, B filter was small.
While the UV emission may be thermal, we model the soft X-ray flux as an excess
on top of a significant power-law component which extends out to at least 10 keV. 
Weak radio emission, peaking at 1.5 GHz, was detected 21 months after discovery.
From the rapid
X-ray variability we calculate the mass of the black hole to be
$1\times10^{6}<M_{BH}<\sim1\times10^{7}$ \msolar 
 and constrain the location of the X-ray emission to within $<$73 $R_{g}$
 of the black hole. The galaxy shows no signs of 
previous AGN activity and we attribute the flare to the accretion
of debris from a star which was tidally destroyed 
during a close approach to the nuclear black hole.
\srcname is perhaps the cleanest example yet found of
a TDE emitting both thermal and non-thermal radiation.
It continues the recently-discovered trend of TDE being 
preferentially hosted in post-starburst galaxies.
If \srcname is a TDE then its flux should continue to reduce over 
the coming years.

\acknowledgements
We thank the anonymous referee and the editor, Sergio Campana, 
for useful comments which
improved the paper, and the XMM OTAC for approving this program.
The XMM-Newton project is an ESA science mission with instruments and contributions directly funded by ESA member states and the USA (NASA).
The \xmm project is supported by the Bundesministerium f\"{u}r Wirtschaft 
und Technologie/Deutches Zentrum f\"{u}r Luft- und Raumfahrt i
(BMWI/DLR, FKZ 50 OX 0001), the Max-Planck Society and the Heidenhain-Stiftung.
We thank the \swift team for approving and performing the monitoring 
observations. This work made use of data supplied by the UK \swift Science Data Centre at 
the University of Leicester.
We are very grateful to Lucas Macri for providing the optical spectrum 
and calibration data taken during the 2MASS redshift survey. 
We thank Pilar Esquej for calculating the
\rosat upper limit, Edo Berger and Giovanni Miniutti for useful discussions 
and Phil Edwards for rapidly scheduling our ATCA observations. The Australia Telescope 
Compact Array is part of the Australia Telescope National Facility which is funded by 
the Australian Government for operation as a National Facility managed by CSIRO.

\section{References}

\bibitem[\protect\citeauthoryear{Alexander et~al.}{2016}]{Alexander16} Alexander, K., Berger, E., Guillochon, J., Zauderer, B., Williams, P. 2016, ApJ, 819, 25
\bibitem[\protect\citeauthoryear{Arcavi et~al.}{2014}]{Arcavi14} Arcavi, I., Gal-Yam, A., Sullivan, M., Pan, Y-C., Cenko, S. et al., 2014, ApJ, 793,38
\bibitem[\protect\citeauthoryear{Bade, Komossa \& Dahlem}{1996}]{Bade96} Bade, N., Komossa, S., Dahlem, M. 1996, A\&A, 309, L35
\bibitem[\protect\citeauthoryear{Bonning et~al.}{2007}]{Bonning07} Bonning, E., Cheng, L., Shields, G., Salviander, S., Gerbhardt, K., 2007, ApJ, 659, 211
\bibitem[\protect\citeauthoryear{Breeveld et~al.}{2010}]{breeveld10} Breeveld, A., Curran, P., Hoversten, E., Koch, S., Landsman, W. et al. 2010, MNRAS, 406, 1687
\bibitem[\protect\citeauthoryear{Bruzual \& Charlot}{2003}]{Bruzual} Bruzual, G. \& Charlot, S. 2003, MNRAS, 344, 1000
\bibitem[\protect\citeauthoryear{Brown et~al.}{2015}]{Brown15} Brown, G., Levan, A., Stanway, E., et al. 2015, MNRAS, 452, 4297
\bibitem[\protect\citeauthoryear{Burrows et~al.}{2005}]{Burrows05} Burrows, D., Hill, J., Nousek, J. et al. 2005, Space Sci. Rev., 120, 165  
\bibitem[\protect\citeauthoryear{Cardelli et~al.}{1989}]{Cardelli} Cardelli, J., Clayton, G., Mathis, J. 1989, ApJ, 345, 245
\bibitem[\protect\citeauthoryear{Cenko et~al.}{2012b}]{Cenko12b} Cenko, S., Krimm, H., Horesh, A. et al., 2012, ApJ, 753, 77
\bibitem[\protect\citeauthoryear{Cenko et~al.}{2016}]{Cenko16} Cenko, S., Cucchiara, A., Roth, N., Veilleux, S., Prochaska, J. et al., 2016, ApJ, 818, 32
\bibitem[\protect\citeauthoryear{Cole et~al.}{2001}]{Cole2001} Cole, S., Norberg, P., Baugh, C., Frenk, C., Bland-Hawthorn, J. et al. 2001, MNRAS 326 255
\bibitem[\protect\citeauthoryear{Condon et~al.}{2002}]{Condon} Condon, J., Cotton, W. \& Broderick, J. 2002, AJ, 123, 188
\bibitem[\protect\citeauthoryear{Done et~al.}{2012}]{Done12} Done, C, Davis, S, Jin, C., Blaes, O., Ward, M., 2012, MNRAS 420, 1848
\bibitem[\protect\citeauthoryear{Donley et~al.}{2002}]{Donley} Donley, J., Brandt, W., Eracleous, M., Boller, Th. 2002, AJ 124, 1308
\bibitem[\protect\citeauthoryear{Edelson et. al.}{2002}]{Edelson02}Edelson, R., Turner, T., Pounds, K, et al., 2002, ApJ, 568, 610
\bibitem[\protect\citeauthoryear{Esquej et~al.}{2007}]{Esquej07} Esquej, P., Saxton, R., Freyberg, M. et al., 2007, A\&A, 462L, 49
\bibitem[\protect\citeauthoryear{Esquej et~al.}{2008}]{Esquej08} Esquej, P., Saxton, R., Komossa, S., Read, A., Freyberg, M. J., 2008, A\&A, 489, 543 
\bibitem[\protect\citeauthoryear{Evans et~al}{2009}]{Evans} Evans, P., Beardmore, A., Page, K. et al. 2009, MNRAS, 397, 1177 
\bibitem[\protect\citeauthoryear{Fabian et~al.}{2002}]{Fabian2002} Fabian, A., Ballantyne D., Merloni, A., Vaughan, S., Iwasawa, K. et~al. 2002, MNRAS, 331, L35 
\bibitem[\protect\citeauthoryear{Cid Fernandes et~al.}{2005}]{Fernand05} Cid Fernandes, R., Mateus, M., Sodre Jr., L., Stasinka, G., Gomes, J. 2005, MNRAS, 358, 363
\bibitem[\protect\citeauthoryear{Cid Fernandes et~al.}{2011}]{Fernand11} Cid Fernandes, R., Stasinka, G., Mateus, A., Vale Asari, N. 2011, MNRAS, 413, 1687
\bibitem[\protect\citeauthoryear{Filippenko}{1982}]{Filippenko} Filippenko, A. 1982, PASP, 94, 715
\bibitem[\protect\citeauthoryear{French et~al.}{2016}]{French} French, K., Arcavi, I., Zabludoff, A. 2016, ApJ, 818, 21
\bibitem[\protect\citeauthoryear{Gabriel et~al.}{2003}]{Gabriel} Gabriel, C. et al. : \emph{}, In ASP Conf. Ser., Vol. 314, ADASS Xiii ed.  Oschenbein, F., Allen, M. \& Egret, D., 759 (2003).
\bibitem[\protect\citeauthoryear{Gehrels et~al.}{2004}]{Gehrels04} Gehrels, N., Chincarini, G., Giommi, P., Mason, K. O., Nousek, J. A. et al. 2004, ApJ, 611, 1005
\bibitem[\protect\citeauthoryear{Gezari et~al.}{2006}]{Gezari06} Gezari, S., Martin, D., Milliard, B. et al., 2006, ApJ, 653L, 25 
\bibitem[\protect\citeauthoryear{Gezari et~al.}{2008}]{Gezari08} Gezari, S., Basa, S., Martin, D. et al., 2008, ApJ, 676, 944
\bibitem[\protect\citeauthoryear{Gezari et~al.}{2009}]{Gezari09} Gezari, S., Heckman, T., Cenko, S. et al., 2009, ApJ, 698, 1367 
\bibitem[\protect\citeauthoryear{Gezari et~al.}{2012}]{Gezari12} Gezari, S., Chornock, R., Rest, A et al., 2012, Nat., 485, 217
\bibitem[\protect\citeauthoryear{Goad et~al.}{2007}]{Goad} Goad, M., Tyler, L., Beardmore, A. et al. 2007, A\&A, 476, 1401
\bibitem[\protect\citeauthoryear{Guillochon, Manukian \& Ramirez-Ruiz}{2013}]{GuillochonPSJ} Guillochon J., Manukian H., Ramirez-Ruiz E., 2013, ApJ, preprint (arXiv:1304.6397)
\bibitem[\protect\citeauthoryear{Halpern, Gezari \& Komossa}{2004}]{Halpern04} Halpern, J., Gezari, S., Komossa, S., 2004, ApJ, 604, 572
\bibitem[\protect\citeauthoryear{Hardt \& Maraschi}{1991}]{HarMar} Hardt, \& Maraschi, 1991
\bibitem[\protect\citeauthoryear{Hills}{1975}]{Hills} Hills, J., 1975, Nat. 254, 295
\bibitem[\protect\citeauthoryear{Holoien et~al.}{2016a}]{Holoien16a} Holoien, T., Kochanek, C., Prieto, J., Stanek, K., Dong, S. et al., 2016, MNRAS, 455, 2918
\bibitem[\protect\citeauthoryear{Holoien et~al.}{2016b}]{Holoien16b} Holoien, T., Kochanek, C., Prieto, J., Grupe, D., Chen, P. et al., 2016, arXiv:1602.01088
\bibitem[\protect\citeauthoryear{Huchra et~al.}{2012}]{Huchra12} Huchra,J., Macri, L., Masters, K., Jarrett, T., Berlind, P. et al. 2012, ApJS, 199, 26
\bibitem[\protect\citeauthoryear{Irwin et~al.}{2015}]{Irwin} Irwin, J., Henriksen, R., Krause, M., Wang, D., Wiegert, T. et al. 2015, ApJ, 809, 172
\bibitem[\protect\citeauthoryear{Jansen et~al.} {2001}]{jansen} Jansen, F., Lumb, D., Altieri, B. et~al 2001. \aap, 365, L1-6
\bibitem[\protect\citeauthoryear{Kara et~al.}{2016}]{Kara16} Kara, E., Miller, J., Reynolds, C., Dai, L. 2016, arXiv:1606.06736
\bibitem[\protect\citeauthoryear{Komossa}{2002}]{Komossa02} Komossa, S., 2002, RvMA 15, 27
\bibitem[\protect\citeauthoryear{Komossa \& Bade}{1999}]{KomossaBade} Komossa, S., Bade, N., 1999, A\&A, 343, 775 
\bibitem[\protect\citeauthoryear{Komossa \& Greiner}{1999}]{Komossa99b} Komossa, S., Greiner, J., 1999, A\&A, 349, L45 
\bibitem[\protect\citeauthoryear{Komossa et~al.}{2004}]{Komossa1242} Komossa, S., Halpern, J., Schartel, N. et al. 2004, ApJ, 603, L17 
\bibitem[\protect\citeauthoryear{Komossa et~al.}{2008}]{Komossa08} Komossa, S., Zhou, H., Wang, T. et al., 2008,  ApJ, 678, 13
\bibitem[\protect\citeauthoryear{Lamastra et~al.}{2009}]{Lamastra09} Lamastra, A., Bianchi, S., Matt, G. et al., 2009, A\&A, 504, 73
\bibitem[\protect\citeauthoryear{Laor}{1991}]{Laor91} Laor, A. 1991, ApJ, 376, 90
\bibitem[\protect\citeauthoryear{Lei et~al.}{2016}]{Lei16} Lei, W-H, Yuan, Q, Zhang, B, Wang, D. 2016, ApJ, 816, 20
\bibitem[\protect\citeauthoryear{Levan et~al.}{2011}]{Levan11} Levan, A., Tanvir, N., Cenko, S. et al.  2011, Sci 333, 199
\bibitem[\protect\citeauthoryear{Lin et~al.}{2015}]{Lin15} Lin, D., Maksym. W., Irwin, J. ,Komossa, S., Webb, N. et al. 2015, ApJ, 811, 43
\bibitem[\protect\citeauthoryear{Maksym et~al.}{2010}]{Maksym10} Maksym, W., Ulmer, M., Eracleous, M.,  2010, ApJ, 722, 1035
\bibitem[\protect\citeauthoryear{Marconi \& Hunt}{2003}]{MarconiHunt} Marconi, A., Hunt, L. 2003, ApJ, 589, L21
\bibitem[\protect\citeauthoryear{Merloni et~al.}{2015}]{Merloni} Merloni, A., Dwelly, T., Salvato, M., Georgakakis, A., Greiner, J., et al, 2015, MNRAS, 452, 69
\bibitem[\protect\citeauthoryear{Miller et~al.}{2015}]{Miller} Miller, J., Kaastra, J., Miller, M., Reynolds, M., Brown, G. et al., 2015, Nat., 526, 542
\bibitem[\protect\citeauthoryear{Moustakas et~al.}{2006}]{Moustakas} Moustakas, J, Kennicutt, R. \& Tremonti, C. 2006, ApJ, 642, 775 
\bibitem[\protect\citeauthoryear{Nandra et~al.}{2007}]{Nandra07} Nandra, K., O'Neill, P., George, I. \& Reeves, J. 2007, MNRAS, 382, 194
\bibitem[\protect\citeauthoryear{Nikolajuk \& Walter}{2013}]{Walter13} Nikolajuk, M.,  Walter, R., 2013, A\&A, 552, 75
\bibitem[\protect\citeauthoryear{Panessa et~al.}{2006}]{Panessa06}Panessa, F., Bassani, L., Cappi, M et al. 2006, A\&A, 455, 173
\bibitem[\protect\citeauthoryear{Phinney}{1989}]{Phinney} Phinney, E.S., 1989, vol 136 of IAU symposium , 543
\bibitem[\protect\citeauthoryear{Ponti et~al.}{2012}]{Ponti} Ponti, G., Papadakis, I., Bianchi, S., Guainazzi, M., Matt, G. et al. 2012, A\&A, 542, A83
\bibitem[\protect\citeauthoryear{Poole et~al.}{2008}]{Poole} Poole, T., Breeveld, A., Page, M. et al. 2008, MNRAS, 383, 627
\bibitem[\protect\citeauthoryear{Read et~al.}{2008}]{Read08} Read, A., Saxton, R., Torres, M. et al.  2008, A\&A, 482, L1
\bibitem[\protect\citeauthoryear{Rees}{1988}]{Rees88} Rees, 1988, Nature, 333, 523 
\bibitem[\protect\citeauthoryear{Reeves et~al.}{2008}]{reeves} Reeves, J., Done, C., Pounds, K. et al. 2008, MNRAS, 385, 108
\bibitem[\protect\citeauthoryear{Roming et~al.}{2005}]{Roming} Roming, P. et al. 2005, SSRv 120, 95
\bibitem[\protect\citeauthoryear{Ross \& Fabian}{2005}]{RossFabian} Ross, R. \& Fabian, A. 2005, MNRAS, 358, 211
\bibitem[\protect\citeauthoryear{Saxton et~al.}{2008}]{Saxton08} Saxton, R., Read, A., Esquej, P. et al.  2008, A\&A 480, 611 
\bibitem[\protect\citeauthoryear{Saxton et~al.}{2012}]{Saxton12} Saxton, R., Read, A., Esquej, P. et al.,  2012, A\&A, 541, 106
\bibitem[\protect\citeauthoryear{Saxton et~al.}{2014}]{Saxton14} Saxton, R., Read, A., Komossa, S., Rodriguez-Pascual, P., Miniutti, G. et al. 2014, A\&A, 572, 1
\bibitem[\protect\citeauthoryear{Schlafly \& Finkbeiner}{2011}]{schlaf} Schlafly, E.  \& Finkbeiner, D. 2011, ApJ, 737, 103
\bibitem[\protect\citeauthoryear{Stern et~al.}{2012}]{Stern12} Stern, D., Assef, R., Benford, D., Blain, A., Cutri, R. et al. 2012, ApJ, 753, 30
\bibitem[\protect\citeauthoryear{Strubbe \& Quataert}{2009}]{Strubbe09} Strubbe, L., Quataert, E., 2009, MNRAS, 400, 2070 
\bibitem[\protect\citeauthoryear{Strubbe \& Quataert}{2011}]{Strubbe11} Strubbe, L.E. \& Quataert, E. 2011, MNRAS, 415, 168
\bibitem[\protect\citeauthoryear{Strubbe \& Murray}{2015}]{Strubbe15} Strubbe, L.E. \& Murray, N. 2015, MNRAS, 454, 2321
\bibitem[\protect\citeauthoryear{Sunyaev \& Titarchuk}{1980}]{SunTit} Sunyaev, R. \& Titarchuk, T. 1980, 
\bibitem[\protect\citeauthoryear{van Velzen et~al.}{2011}]{vanVelzen11} van Velzen, S., Farrar, G., Gezari, S. et al., 2011, ApJ, 741, 73
\bibitem[\protect\citeauthoryear{van Velzen \& Farrar}{2014}]{vvFar14} van Velzen, S., \& Farrar, G. 2014, ApJ, 792, 53
\bibitem[\protect\citeauthoryear{van Velzen et~al.}{2016}]{vanVelzen16} van Velzen, S., Anderson, G., Stone, N., Fraser, M., Wevers, T. et al. 2016, Sci, 351, 62 
\bibitem[\protect\citeauthoryear{Vaughan}{2003}]{Vaughan03} Vaughan, S., Edelson, R., Warwick, R., Uttley, P. 2003, MNRAS, 345, 1271 
\bibitem[\protect\citeauthoryear{Vaughan, Edelson \& Warwick}{2004}]{Vaughan04} Vaughan, S., Edelson, R., \& Warwick, R. 2004, MNRAS, 349, L1 
\bibitem[\protect\citeauthoryear{Wang et~al.}{2011}]{Wang11} Wang, T.-G., Zhou, H.-Y., Wang, L.-F., Lu, H.-L., Xu, D., 2011, ApJ, 740, 85
\bibitem[\protect\citeauthoryear{Wilkins et~al.}{2012}]{Wilkins12} Wilkins, S., Gonzalez-Perez, V., Lacey, C., Baugh, C. 2012, MNRAS, 424, 1522
\bibitem[\protect\citeauthoryear{Willingale et~al.}{2013}]{Willingale} Willingale, R., Starling, R., Beardmore, A., Tanvir, N., O'Brien, P., 2013, MNRAS, 431, 394
\bibitem[\protect\citeauthoryear{Wilms, Allen \& McCray}{2000}]{Wilms} Wilms, J., Allen, A., \& McCray, R. 2000, ApJ 542, 914
\bibitem[\protect\citeauthoryear{Zauderer et~al.}{2011}]{Zauderer11} Zauderer, B., Berger, E., Soderberg, A. et al. 2011, Nat. 476, 425

%\bibliographystyle{abbrvnat}
%\bibliographystyle{unsrtnat}
%\section{References}

%\bibliographystyle{abbrvnat}
%\bibliographystyle{unsrtnat}
%\bibliographystyle{plainnat}
%\bibliography{references}{}

\end{document}